\documentclass[tighten,nolinenumbers, notrackchanges, twocolumn]{aastex62m}

\usepackage{amsmath}
\usepackage{amssymb}
\usepackage{amssymb}
\usepackage{mathtools}
\usepackage{mathrsfs}

\usepackage{morefloats}
\usepackage{longtable}
\usepackage{afterpage}
\usepackage{float}

\hypersetup{linkcolor=red,citecolor=blue,urlcolor=magenta}

\usepackage{soul}

\usepackage[makeroom]{cancel}

\graphicspath{{figs/}}

\usepackage{xspace}

\newcommand*{\XPSI}{X-PSI\xspace}
\newcommand*{\NICER}{NICER\xspace}
\newcommand*{\xmm}{XMM-Newton\xspace}
\newcommand*{\MultiNest}{\textsc{MultiNest}\xspace}

\newcommand{\msol}{$M_\odot$\xspace}
\newcommand{\jdbl}{PSR~J0030$+$0451\xspace}
\newcommand{\joh}{PSR~J0740$+$6620\xspace}

\newcommand{\TT}[1]{\texttt{#1}}

\newcommand{\be}{\begin{equation}}
\newcommand{\ee}{\end{equation}}

\defcitealias{Riley2021}{R21}
\defcitealias{Wolff20}{W21}

\received{2022 July 20}
\revised{2022 September 11}
\accepted{2022 September 23}
\published{2022 December 19}

\publishedjournal{The Astrophysical Journal} 

\shorttitle{Radius of PSR J0740+6620}
\shortauthors{Salmi~et~al.}

\begin{document}

\title{The Radius of PSR J0740+6620 from NICER with NICER Background Estimates}

\correspondingauthor{T.~Salmi}
\email{t.h.j.salmi@uva.nl}
  
\author[0000-0001-6356-125X ]{Tuomo~Salmi}
\affil{Anton Pannekoek Institute for Astronomy, University of Amsterdam, Science Park 904, 1090GE Amsterdam, the Netherlands}

\author[0000-0003-3068-6974]{Serena~Vinciguerra}
\affil{Anton Pannekoek Institute for Astronomy, University of Amsterdam, Science Park 904, 1090GE Amsterdam, the Netherlands}

\author[0000-0002-2651-5286]{Devarshi~Choudhury}
\affil{Anton Pannekoek Institute for Astronomy, University of Amsterdam, Science Park 904, 1090GE Amsterdam, the Netherlands}

\author[0000-0001-9313-0493]{Thomas~E.~Riley}
\affil{Anton Pannekoek Institute for Astronomy, University of Amsterdam, Science Park 904, 1090GE Amsterdam, the Netherlands}

\author[0000-0002-1009-2354]{Anna~L.~Watts}
\affil{Anton Pannekoek Institute for Astronomy, University of Amsterdam, Science Park 904, 1090GE Amsterdam, the Netherlands}

\author[0000-0003-4815-0481]{Ronald A. Remillard}
\affil{MIT Kavli Institute for Astrophysics \& Space Research, MIT, 70 Vassar Street, Cambridge, MA 02139, USA}

\author[0000-0002-5297-5278]{Paul~S.~Ray}
\affil{Space Science Division, U.S. Naval Research Laboratory, Washington, DC 20375, USA}

\author[0000-0002-9870-2742]{Slavko~Bogdanov} 
\affil{Columbia Astrophysics Laboratory, Columbia University, 550 West 120th Street, New York, NY 10027, USA}

\author[0000-0002-6449-106X]{Sebastien~Guillot}
\affil{Institut de Recherche en Astrophysique et Plan\'{e}tologie, UPS-OMP, CNRS, CNES, 9 avenue du Colonel Roche, BP 44346, F-31028 Toulouse Cedex 4, France}

\author{Zaven~Arzoumanian}
\affil{X-Ray Astrophysics Laboratory, NASA Goddard Space Flight Center, Code 662, Greenbelt, MD 20771, USA}

\author[0000-0003-2759-1368]{Cecilia Chirenti}
\affil{Department of Astronomy and Joint Space-Science Institute, University of Maryland, College Park, MD 20742-2421, USA}
\affil{Astroparticle Physics Laboratory NASA/GSFC, Greenbelt, MD 20771, USA}
\affil{Center for Research and Exploration in Space Science and Technology, NASA/GSFC, Greenbelt, MD 20771, USA}
\affil{Center for Mathematics, Computation, and Cognition, UFABC, Santo Andr\'{e}, SP 09210-170, Brazil} 

\author[0000-0001-6157-6722]{Alexander~J.~Dittmann}
\affil{Department of Astronomy and Joint Space-Science Institute, University of Maryland, College Park, MD 20742-2421, USA}
\affil{Theoretical Division, Los Alamos National Laboratory, Los Alamos, NM 87545, USA}

\author[0000-0001-7115-2819]{Keith~C.~Gendreau}
\affil{X-Ray Astrophysics Laboratory, NASA Goddard Space Flight Center, Code 662, Greenbelt, MD 20771, USA}

\author[0000-0002-6089-6836]{Wynn~C.~G.~Ho}
\affil{Department of Physics and Astronomy, Haverford College, 370 Lancaster Avenue, Haverford, PA 19041, USA}

\author[0000-0002-2666-728X]{M.~Coleman~Miller}
\affil{Department of Astronomy and Joint Space-Science Institute, University of Maryland, College Park, MD 20742-2421, USA}

\author[0000-0003-4357-0575]{Sharon~M.~Morsink}
\affil{Department of Physics, University of Alberta, 4-183 CCIS, Edmonton, AB, T6G 2E1, Canada}

\author[0000-0002-9249-0515]{Zorawar~Wadiasingh}
\affiliation{Department of Astronomy, University of Maryland, College Park, Maryland 20742, USA}
\affiliation{Astrophysics Science Division, NASA Goddard Space Flight Center, Greenbelt, MD 20771, USA.}
\affiliation{Center for Research and Exploration in Space Science and Technology, NASA/GSFC, Greenbelt, MD 20771, USA}

\author[0000-0002-4013-5650]{Michael~T.~Wolff}
\affil{Space Science Division, U.S. Naval Research Laboratory, Washington, DC 20375, USA}

\begin{abstract}

We report a revised analysis for the radius, mass, and hot surface regions of the massive millisecond pulsar PSR J0740+6620, studied previously with joint fits to NICER and \xmm data by Riley et al. (2021) and Miller et al. (2021). We perform a similar Bayesian estimation for the pulse-profile model parameters, except that instead of fitting simultaneously the \xmm data, we use the best available NICER background estimates to constrain the number of photons detected from the source. This approach eliminates any potential issues in the cross-calibration between these two instruments, providing thus an independent check of the robustness of the analysis. 
The obtained neutron star parameter constraints are compatible with the already published results, 
with a slight dependence on how conservative the imposed background limits are. A tighter lower limit causes the inferred radius to increase, and a tighter upper limit causes it to decrease.
We also extend the study of the inferred emission geometry to examine the degree of deviation from antipodality of the hot regions. 
We show that there is a significant offset to an antipodal spot configuration, mainly due to the non-half-cycle azimuthal separation of the two emitting spots. 
The offset angle from the antipode is inferred to be above $25\degr$ with 84\% probability. 
This seems to exclude a centered-dipolar magnetic field in PSR J0740+6620. 
\end{abstract}

\section{Introduction}
\label{sec:intro}

Determination of the masses and radii of a set of neutron stars (NSs) provides direct insight into the equation of state (EOS) of matter at supranuclear densities \citep{Lattimer12ARNPS,OE17,Baym2018,Tolos2020,YangPiek2020,Hebeler2021}. 
This is the primary science goal of NASA's Neutron Star Interior Composition Explorer 
\citep[\NICER;][]{Gendreau2016}, an X-ray telescope installed on the International Space Station (ISS). 
\NICER aims to measure masses and radii of rotation-powered millisecond X-ray pulsars using the technique of pulse-profile modeling \citep[PPM; see][and references therein]{2016RvMP...88b1001W,Watts2019,Bogdanov2019b,Bogdanov2021}. 
PPM exploits relativistic effects that leave a measurable signature on rotationally modulated emission from the hot ($\sim 10^6$ K) magnetic polar caps of the pulsars. 
By modeling the relativistic effects, such as gravitational light-bending, Doppler boosting, aberration, time delays, and the effects of rotationally induced stellar oblateness, we are able to infer not only the mass and radius but also the geometric surface properties of the pulsar, in effect making a map of the hot emitting regions. 
Those regions are heated by the flow-back of particles that are accelerated in the magnetosphere. 
The bombarding particles are likely ultra-relativistic and produced in single photon magnetic pair cascades, originating at the open field line region where field lines connect the pulsar to its light cylinder \citep[see, e.g.,][]{RudermanSutherland1975,arons81,HM01}. 

To date the NICER collaboration have reported results for two pulsars: \jdbl \citep{MLD_nicer19,Riley2019} and the high-mass pulsar \joh \citep[][hereafter R21]{Miller2021,Riley2021}.\footnote{Each source is analyzed by two independent teams within the collaboration, each with their own PPM code and analysis pipeline.} The results have pointed to non-antipodal magnetic polar cap geometries \citep[suggesting a complex magnetic field structure;][]{Bilous_2019,MLD_nicer19,Chen2020,Kalapotharakos2021}, and are starting to restrict dense matter models \citep[for a small selection of recent papers, see for example][]{Legred21,JieLiJ21,Miller2021,Pang21,Raaijmakers2021,TangSP21,Annala2022,Biswas2022}.

The pulse profiles are formed by recording the energy and arrival times by the individual \NICER Focal Plane Modules (FPMs), and rotation phases are assigned via a previously observed pulsar ephemeris (obtained from radio timing) for every detected event.
We then sum the total number of events in each energy and phase bin \citepalias[see for example Figure 1 of][]{Riley2021}. 
Because the \NICER pulsar sources are assumed to be extremely stable rotators, pulse profiles can be built up over weeks or even years of observations, in order to obtain the required high signal-to-noise ratio. 

The resulting pulse profile consists of pulsed and unpulsed components. The hot spots can produce both pulsed and unpulsed emission, with unpulsed emission likely generated if a spot overlaps the rotational pole, if the viewing angle is close to the rotational pole, or if the compactness is high enough and the geometry is such that parts of the spots remain in view as the star rotates. 
The unpulsed component also has background contributions from instrumental noise, particles, the cosmic X-ray background, and other astrophysical sources in the field of view (FoV) \footnote{\NICER has a 6' non-imaging FoV.}. 
It is crucial for the NS parameter estimation to accurately determine the relative contributions to the unpulsed emission from the background and the hot spots. 
If, for example, the background is underpredicted, the hot spots will be incorrectly assigned a larger unpulsed component, possibly leading to an incorrect prediction of a larger gravitational lensing effect caused by a larger compactness ratio, $M/R$. 
This issue is most important for very faint pulsars with a pulsed signal that is only slightly larger than the background, such as \joh.

Having an independent constraint on the unpulsed background thus provides useful input to the inference analysis.  At the time of the 2019 analysis of \jdbl \citep{MLD_nicer19,Riley2019}, \NICER background models were not sufficiently well developed to be used in the inference process. Instead, the inferred \NICER phase-averaged source spectrum was compared a posteriori to a phase-averaged source spectrum derived independently from \xmm observations (for which the background is empirically determined and better constrained from the observations, because \xmm focuses photons and images the region of interest). 
These results indicated that the \NICER PPM analysis might be attributing slightly too much of the unpulsed component to the non-source background (see Figure 15 of \citealt{Riley2019}; and Figure 12 of \citealt{Miller2021}). 

In the 2021 analysis of \joh, a different approach was taken: we performed simultaneous inference of both the phase-resolved \NICER data set and a smaller phase-averaged \xmm data set, using a blank-sky estimate of the astrophysical background as an additional constraint on the \xmm fit.  This was an indirect method of constraining the non-source contribution to the unpulsed component in the \NICER pulse profile. However it had a large effect, shifting the median posterior radius upwards by $\sim 1$ km (compared to an analysis using \NICER data alone), by excluding higher compactness solutions \citetext{\citealt{Miller2021}; \citetalias{Riley2021}}. The inferred \NICER background was compared a posteriori to the \NICER background model estimates that existed at the time, and was found to be in reasonable agreement 
\citepalias[see Figure 15 of][]{Riley2021}.   

One disadvantage of using \xmm to provide a background constraint is that one has to take into account both the absolute uncertainty in X-ray flux calibration and the cross-calibration uncertainty from instrument to instrument. 
While we made an attempt to capture this in our previous analysis by including energy-independent calibration model parameters, we did not take into account any potential energy-dependent cross-calibration effects.

Thanks to intensive work by the \NICER instrument team, 
there are now reliable \NICER background estimates that can be used directly in the PPM inference pipeline to constrain the non-source contribution to the unpulsed component in the pulse profile.  
In this paper, we update the \citetalias{Riley2021} analysis of \joh, using these constraints as an alternative to relying on \xmm.  
We apply both the space weather (SW) background estimate \citep{SpaceWeather} to the original \NICER data from \citet{Wolff20}, and the 3C50 background estimate using a new 3C50-filtered data set \citep{remi22}. 
We find that the results are consistent with those presented in \citetalias{Riley2021}, and explore and refine the procedure to be used for \NICER background estimates in future analyses.
In Appendix \ref{apndx:A}, an independent analysis is presented using the 3C50 data set based on the codes and methods from \citet{Miller2021}.

The remainder of this paper is structured as follows. 
In Section \ref{sec:data_and_bkg}, we introduce the different data sets and background estimates used for \joh. 
In Section \ref{sec:methods}, we present the modeling procedure to obtain NS parameter constraints using PPM, with the help of those data and estimates.   
In Section \ref{sec:results}, we show the results and compare them to the previously published NS parameter constraints. 
We also analyze the spot geometry of the best-fitting samples and perform more simulations to quantify the likely deviation from an antipodal spot configuration.  
We discuss the implications arising from our results in Section \ref{sec:discussion} and conclude in Section \ref{sec:conclusions}.

\section{Data Analysis and Background Modeling}\label{sec:data_and_bkg}

In this section, we summarize the three event data sets and background estimates used in the analysis presented in this paper.

\subsection{Previous Data Set with Space Weather Background Estimate}\label{sec:model_sw}

We start by using the same X-ray event data described in \citet{Wolff20} and used by \citetalias{Riley2021} and \citet{Miller2021}, which we will refer to as the W21 data set. This filtered data set has 1.60268 Ms of exposure accumulated over the period 2018 September 21 through 2020 April 17. 
The data set contains thousands of individual good time intervals (GTIs), which are filtered based on the sorting method of \citet{Guillot2019} without any additional GTI consolidation. We note that \citet{Essick2022} raised questions about the GTI sorting method. 
We investigated and found that using this GTI sorting method for \joh is not an issue (see \citealt{Essick2022}). The idealized model considered in \citet{Essick2022} is a poor representation of NICER data for which the non-pulsar background contributes the majority of accumulated counts and is not constant. This means that, in practice, the GTI sorting procedure does not significantly alter the count-rate distribution of the pulsed emission in the data set analyzed in \citet{Miller2021}, \citetalias{Riley2021} and, as a consequence, does not introduce a measurable bias in the inferred NS radius.

For this work, we have also generated a background estimate using the SW background estimator \citep{SpaceWeather}.
This estimator is based only on environmental conditions when the data were taken. The parameters are the cosmic ray cutoff rigidity (the COR\_SAX parameter) and the planetary K index ($K_p$, characterizing the magnitude of geomagnetic storms). The estimated background for an observation is constructed using spectra from blank-sky observations obtained within the same ranges of those two parameters. When combined with the COR\_SAX$>2$ cut\footnote{Note that there is an error in \citetalias{Riley2021} where it incorrectly specifies the filter criterion as COR\_SAX$>5$.} used to create this data set (which excludes the regions of highest and most-variable background), this method has been shown to be very reliable. When averaged over many GTIs, a background lower limit of 0.9 times the SW estimate is believed to be a safe prior to impose in the modeling. 
We note that this choice is based on extensive experience applying the SW model to a larger number of data sets, while it remains subject to more-rigorous study of the systematic errors (overprediction or underprediction) in the SW model. 
Other assumptions for the lower limit can affect the inferred results, as mentioned in Section \ref{sec:nicer_SW}. 

The count-rate spectrum for the W21 data set and the SW background are shown in Figure \ref{fig:W21_vs_3C50_data} in comparison to the corresponding 3C50 data and background estimate, which are introduced next.

{
    \begin{figure}[t!]
    \centering
    \resizebox{\hsize}{!}{\includegraphics[
    width=\textwidth]{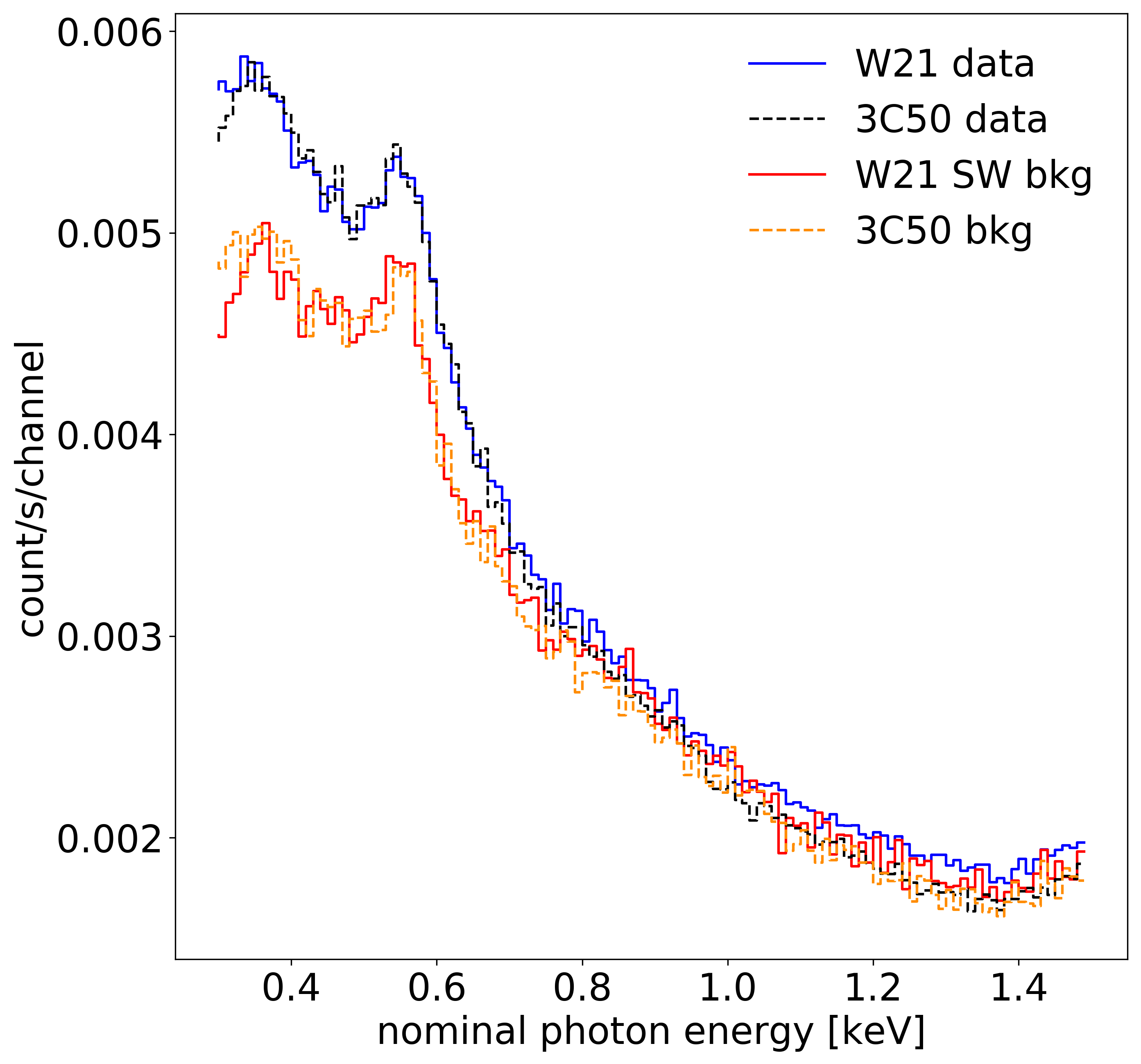}}
    \caption{\small{
    \NICER count-rate spectrum of \joh as a function of detector channel nominal photon energy  (i.e., the PI value from the event file divided by 100).    
    The blue solid and the black dashed step functions represent the spectrum for W21 and 3C50 data sets, respectively. 
    The former is the same as used in \citetalias{Riley2021}, \citet{Miller2021}. 
    The space weather background estimate (specific for the W21 data) is shown with a red solid step function, and the 3C50 background estimate (specific for the 3C50 data) is shown with an orange dashed step function.}
    }
    \label{fig:W21_vs_3C50_data}
    \end{figure}
}

\subsection{3C50 Data and Background Model}\label{sec:3C50_data_and_bkg_maintext}

An alternative, empirical model to estimate the \NICER background, known as the 3C50 model \citep{remi22}, uses three non-source count rates within the on-source GTIs to select and rescale background components from the model's libraries. 
The full description of the 3C50 data and background model used in this work is presented in Appendix \ref{sec:3C50_data_and_bkg}.

\subsection{\xmm EPIC}

As a part of our analysis, we use the same \xmm data as in \citetalias[][]{Riley2021} (see Section 2.1.2. of that paper), pre-processed correspondingly. 
The three EPIC instruments (pn, MOS1, and MOS2) were employed to get phase-averaged spectral information, with effective exposure times of 6.81, 17.96, and 18.7 ks, respectively. 
In addition, blank-sky observations were used to constrain the background contribution in the \xmm data, as in \citetalias[][]{Riley2021}; see Section 2.5.2. of that paper.

\section{Modeling Procedure}\label{sec:methods}

The modeling procedure is largely shared with that of \citet{Riley2019}, \citet{Bogdanov2019b}, \citetalias{Riley2021}, and \citet{Bogdanov2021}. 
In this section, we summarize that methodology and report the new model components in more detail.
For all posterior computations, we use the X-ray Pulse Simulation and Inference (\XPSI) code, version \texttt{v0.7.10} \citep{xpsi}, which belongs to the same \texttt{v0.7} framework as used by \citetalias{Riley2021}\footnote{\url{https://github.com/xpsi-group/xpsi}}. 
The analysis files may be found in the persistent repository on Zenodo\footnote{doi:\href{https://doi.org/10.5281/zenodo.6827537}{10.5281/zenodo.6827537}} 
including the data products, the numeric model files, scripts, and Jupyter notebooks in the Python language to produce the results and figures of this article using the \XPSI framework, and posterior sample files.

\subsection{Radiation from Source to Telescope}

We model the energy-resolved X-ray pulses using the PPM technique with the ``Oblate Schwarzschild'' approximation \citep[see e.g.,][]{ML98,NS02,PG03,MLC07,lomiller13,AGM14,Bogdanov2019b,Watts2019}. 
The choices for prior probability density functions (PDFs) of the model parameters follow those presented in \citetalias{Riley2021}. 
Mass, inclination, and distance priors we obtain from the generalized least-squares (GLS) algorithm determining timing model parameters from wide-band radio timing data \citep{Fonseca20}.
These priors were also used for the production analysis in \citetalias{Riley2021}, although not in the exploratory analysis shown, e.g., in Figure 5 of that paper. 
For relativistic ray-tracing, multiple imaging was still allowed to enable exploration of high compactness solutions due to the very high mass known from the radio timing ($2.08 \pm 0.07$ \msol from \citealt{Fonseca20}). 
The interstellar attenuation of X-rays is again accounted for by neutral hydrogen column density $N_{\rm H}$, based on the state-of-the-art \texttt{TBabs} model \citep[updated in 2016]{Wilms2000}, and having a conservative prior of $N_{\rm H}\sim U(0,10^{21})$~cm$^{-2}$.

\subsection{Surface Hot Regions}\label{sec:surf_hot_regions}

The surface hot regions are modeled using similar methods to \citetalias{Riley2021}, our main model being again \texttt{ST-U} (\textit{Single-Temperature-Unshared}), with two circular hot regions with uniform effective temperature of the atmosphere \citep{Riley2019}. 
In this model, the spots are not restricted to an antipodal reflection symmetry. 
However, to further study the deviation from antipodal symmetry, and to check how much worse antipodal models would perform, we also consider antipodal models (see Section \ref{sec:nicer_STS}).
In this examination, we use both the model \texttt{ST-S} (\textit{Single-Temperature-Shared}) from \citet{Riley2019} where both hot spots share the same temperature and spot size, and a new model version \texttt{ST-Ua} (\textit{Single-Temperature-Unshared-antipodal}) where the spot parameters (size, temperature) are unshared but the spot centers are fixed to be antipodal.
For the priors of the hot region parameters, we follow \citetalias{Riley2021} and use a flat PDF of the cosine of each hot region center colatitude \citep[unlike in][]{Riley2019}.

For the atmosphere model, determining the specific intensity emitted from the NS surface, we restrict ourselves to using only the fully ionized hydrogen model \texttt{NSX} \citep{Ho01}. 
The sensitivity to atmosphere composition, using helium instead of hydrogen, was shown to be negligible in the radius of PSR~J0740$+$6620 for the models and data considered in \citetalias{Riley2021}, although with some small changes in the properties of the hot regions. 
Other uncertainties related to the atmosphere model choice are discussed and treated in \citet{Bogdanov2021} and in T. Salmi et al. (2023, in preparation). 
As in \citetalias{Riley2021}, we assume that the surface exterior to the hot regions is not radiating. 

\subsection{Instrument Response Models}\label{sec:instru_response}

For the \NICER W21 data set, we use the instrument response files described in Section 2.4 of \citetalias{Riley2021}. 
For the 3C50 GTIs, a singular response file is used, and it is calculated as the sum of the individual responses for the 50 FPMs selected for that data set.

For the joint analysis with \xmm, we use the restricted effective-area uncertainty of \citetalias{Riley2021}.  
The overall scaling factor (for each instrument) is defined as a product of a shared scaling factor and a priori statistically independent telescope-specific scaling factor. 
For the restricted uncertainty, a standard deviation of 3\% \footnote{See \href {https://heasarc.gsfc.nasa.gov/docs/heasarc/caldb/nicer/docs/xti/NICER-xti20200722-Release-Notesb.pdf}{heasarc.gsfc.nasa.gov/docs/heasarc/caldb/nicer/docs/xti/
NICER-xti20200722-Release-Notesb.pdf} and \href {https://xmmweb.esac.esa.int/docs/documents/CAL-TN-0018.pdf}{xmmweb.esac.esa.int/docs/
documents/CAL-TN-0018.pdf} for further details.} was assumed for each telescope-specific scaling factor and 10\% for the shared scaling factor \citep{Ishida2011,Madsen2017,Plucinsky2017}, resulting in about 10.4\% uncertainty in the overall scaling factor as in Section 4.2 of \citetalias{Riley2021}. 
We have also explored the sensitivity to the cross-calibration uncertainty by performing low-resolution runs with two other choices for uncertainties, which are shown in Section \ref{sec:nicer_3C50_bkg_XMM}. 
The least restricted uncertainty corresponds to the initial scaling of \citetalias{Riley2021} with $10\%$ uncertainty in both telescope-specific and shared scaling factors, leading to the 15\% uncertainty in the overall scaling factors. 
The most restricted case (more than in Section 4.2 of \citetalias{Riley2021}) applies $5\%$ shared scaling factor uncertainty and $3\%$ telescope-specific uncertainty leading to $5.8 \%$ uncertainty in the overall scaling factors.
For the \NICER analysis without \xmm, we assume $\pm 15 \%$ uncertainty in the effective area of the instrument, by using a free energy-independent effective-area scaling parameter as in  \citetalias{Riley2021}.
We checked (for a few of our runs) that importance-sampling the scaling parameter prior from $\pm 15 \%$ to $\pm 10.4 \%$ has only negligible effects on the posteriors of the other parameters. 
For example, the credible interval for radius remains the same with better than 0.1 \% accuracy. 

\subsection{Likelihood Function and Background}\label{sec:likelihood_background}

In this section, we present how the likelihood function is calculated using different constraints for the background. 
We consider four different cases, which are summarized in Table 
\ref{table:cases}. 

\begin{deluxetable}{cccc}[b]
\tablecaption{Summary of the Model Cases}
\tablehead{\colhead{Case} & \colhead{\NICER $^{\mathrm{a}}$} & \colhead{\NICER+BKG $^{\mathrm{b}}$} & \colhead{XMM+BKG} $^{\mathrm{c}}$} 
\startdata
1 & YES & NO & NO \\
\hline
2 & YES & NO & YES \\
\hline
3 & NO & YES & NO \\
\hline
4 & NO & YES & YES \\
\enddata
\tablecomments{\ \\  $^{\mathrm{a}}$ Analysis without any prior \NICER background information. \\ $^{\mathrm{b}}$ NICER analysis where \NICER background (BKG) estimates are applied. \\ $^{\mathrm{c}}$ Analysis where \xmm data with \xmm background information are applied.}
\end{deluxetable}\label{table:cases}

The first and second cases, when applied to the W21 data set, are identical to those applied in \citetalias{Riley2021}. 
We apply the first and second cases also to the 3C50 data set using the same methodology.
In these cases, the uncertainty in the \NICER background is numerically marginalized by integrating the likelihood over statistically independent background variables $\{\mathbb{E}[b_{\rm N}]\}$ that are allowed to range from $\{0\}$ to an unspecified set of upper bounds $\{\mathcal{U}\}$ \cite[see Appendix B.2 in][for details]{riley_thesis}. 
Here $b_{\rm N}$ denotes the background count rate at a specific \NICER energy channel. 
For the second case when \xmm data is applied, specified limits are used for the \xmm background variables $\{\mathbb{E}[b_{\rm X}]\}$, where $b_{\rm X}$ is the background count rate at a specific \xmm energy channel. 
The lower and upper bounds for the support are $\mathscr{L}\coloneqq\textrm{max}\left(0,\mathscr{B}_{\rm X}-n\sqrt{\mathscr{B}_{\rm X}}\right)$ and $\mathscr{U}\coloneqq\mathscr{B}_{\rm X}+n\sqrt{\mathscr{B}_{\rm X}}$, where $\{\mathscr{B}_{\rm X}\}$ is the set of  \xmm background count numbers based on the blank-sky observation, and $n=4$ is the chosen degree of conservatism.
These blank-sky count numbers are rescaled with the ratio of the areas of the extraction regions of the blank-sky and \joh observation, and divided by the exposure time of the blank-sky observation, when converting them to limits for \xmm background variables $\{\mathbb{E}[b_{\rm X}]\}$, as shown in Equation (3) of \citetalias{Riley2021}. 
Note that in \citetalias{Riley2021} and the associated code and results repository \citep{J0740zenodo} this scaling factor (known also as BACK\_SCAL factor) was the inverse of the correct value.  
Since the scaling factor is close to unity for \joh observations, the effect was anticipated to be minimal, but to be sure, we confirmed this with an additional inference run. 
The results of this additional run, together with corrected scripts, can be found in an update to the repository \citep{J0740zenodo_updated}. 

The background-marginalized likelihood function (probability of the data given the model) for the \NICER and \xmm data sets, assuming a flat prior density in the background variables, is then given as
\begin{align}
\begin{split}
    & p(d_{\rm N}, d_{\rm X}, \{\mathscr{B}_{\rm X}\} \,|\, s)
    \propto
    \mathop{\int}_{\{0\}}^{\{\mathcal{U}\}}
    p(d_{\rm N} \,|\, s, \{\mathbb{E}[b_{\rm N}]\}, \textsc{nicer}) \\
    & \times
    d\{\mathbb{E}[b_{\rm N}]\}
    \mathop{\int}_{\{\mathscr{L}\}}^{\{\mathscr{U}\}}
    p(d_{\rm X} \,|\, s, \{\mathbb{E}[b_{\rm X}]\}, \textsc{xmm})d\{\mathbb{E}[\mathscr{B}_{\rm X}]\},
    \label{eqn:background-marginalized likelihood}
\end{split}
\end{align}
where $d_{\rm X}$ is the \xmm count matrix data, $d_{\rm N}$ is the \NICER count matrix data, $s$ denotes the pulsar signal parameters, \textsc{nicer} and \textsc{xmm} denote models for the response of the corresponding instrument to incident radiation, and $d$'s followed by brackets show the differentials of the integrals.

{
    \begin{figure}[t!]
    \centering
    \resizebox{\hsize}{!}{\includegraphics[
    width=\textwidth]{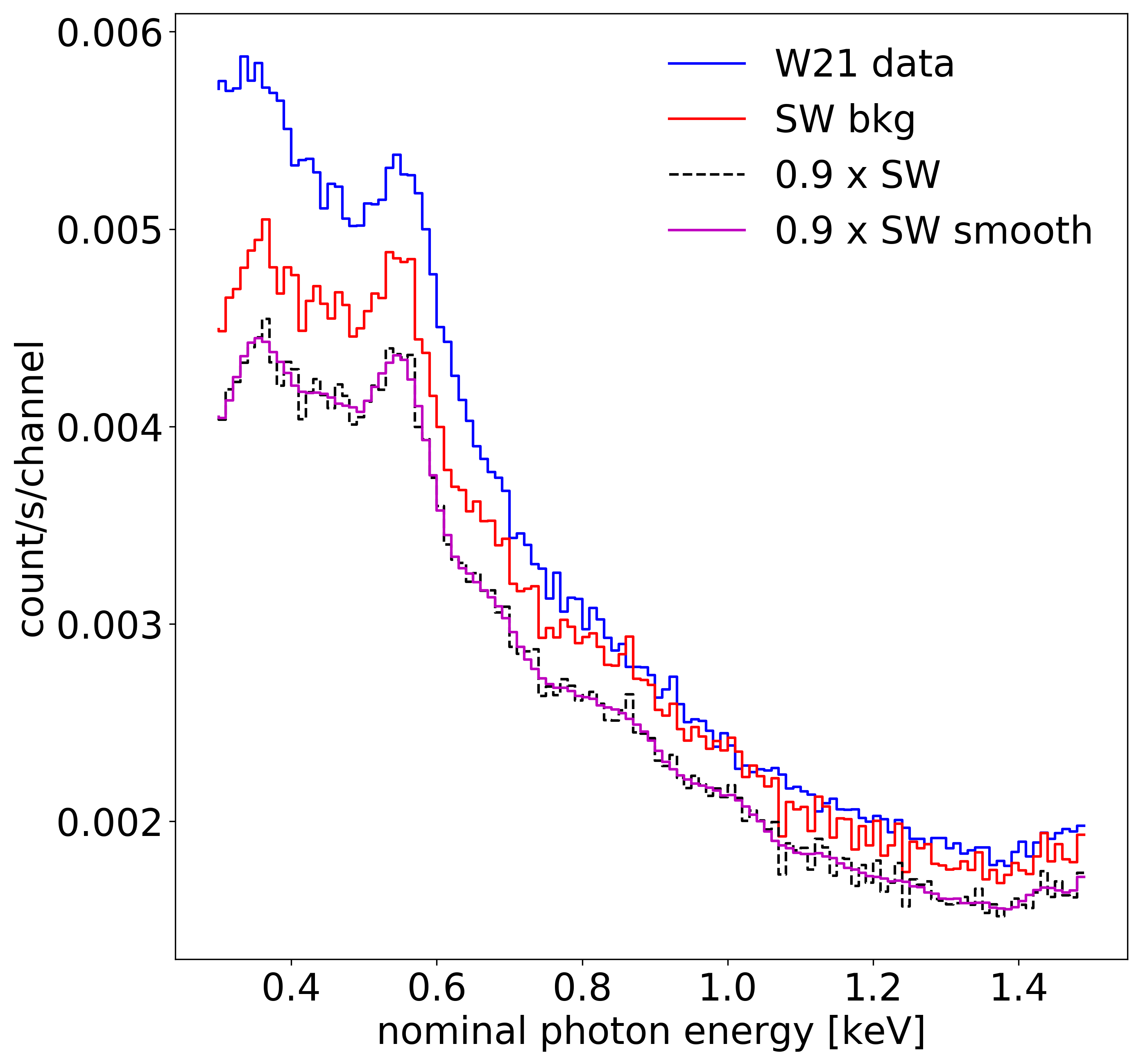}}
    \caption{\small{
    Prior limits for the background of the W21 data set.
    The count-rate spectrum and the SW background estimate are shown with blue and red step functions as in Figure \ref{fig:W21_vs_3C50_data}. 
    The black dashed step function is the estimate multiplied with a factor of 0.9, and a smoothed version of that is shown with a magenta step function. 
    These two rescaled estimates were used as lower limits $\{\mathfrak{L}\}$ for the joint prior PDF of the expected count-rate variables $\{\mathbb{E}[b_{\rm N}]\}$ (see Equation \ref{eqn:background-marginalized_likelihood_new}). 
    }}
    \label{fig:sw}
    \end{figure}
}





{
    \begin{figure}[t!]
    \centering
    \resizebox{\hsize}{!}{\includegraphics[
    width=\textwidth]{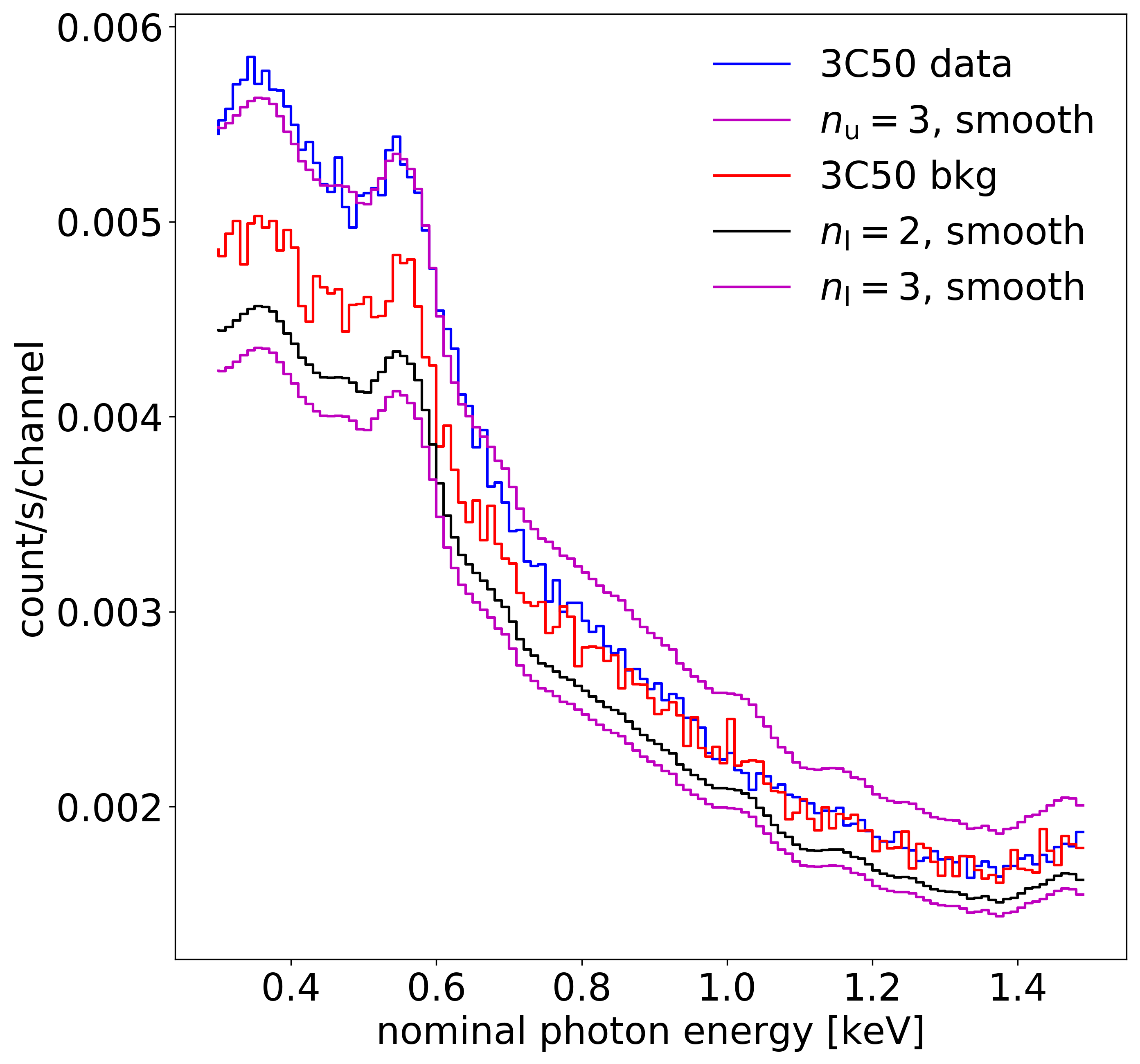}}
    \caption{\small{
    Prior limits for the background of the 3C50 data set. 
    The count-rate spectrum and the 3C50 background estimate are shown with blue and red step functions (corresponding to the black and gray dashed functions in Figure \ref{fig:W21_vs_3C50_data}). 
    The magenta step functions show the smoothed joint prior PDF lower and upper limits for the expected count-rate variables $\{\mathbb{E}[b_{\rm N}]\}$ based on $3 \sigma$ uncertainty (see the equations and main text in Section \ref{sec:likelihood_background}). 
    The black step function shows the corresponding $2 \sigma$ lower limit.
    The corresponding figure with $2 \sigma$ and $3 \sigma$ lower limits based on the minimum of the data and the background estimate (mdb) is shown in the online journal figure set (HTML version). 
    }
    }
    \label{fig:3c50_bkgs}
    \end{figure}
}







{
    \begin{figure}[t!]
    \centering
    \resizebox{\hsize}{!}{\includegraphics[
    width=\textwidth]{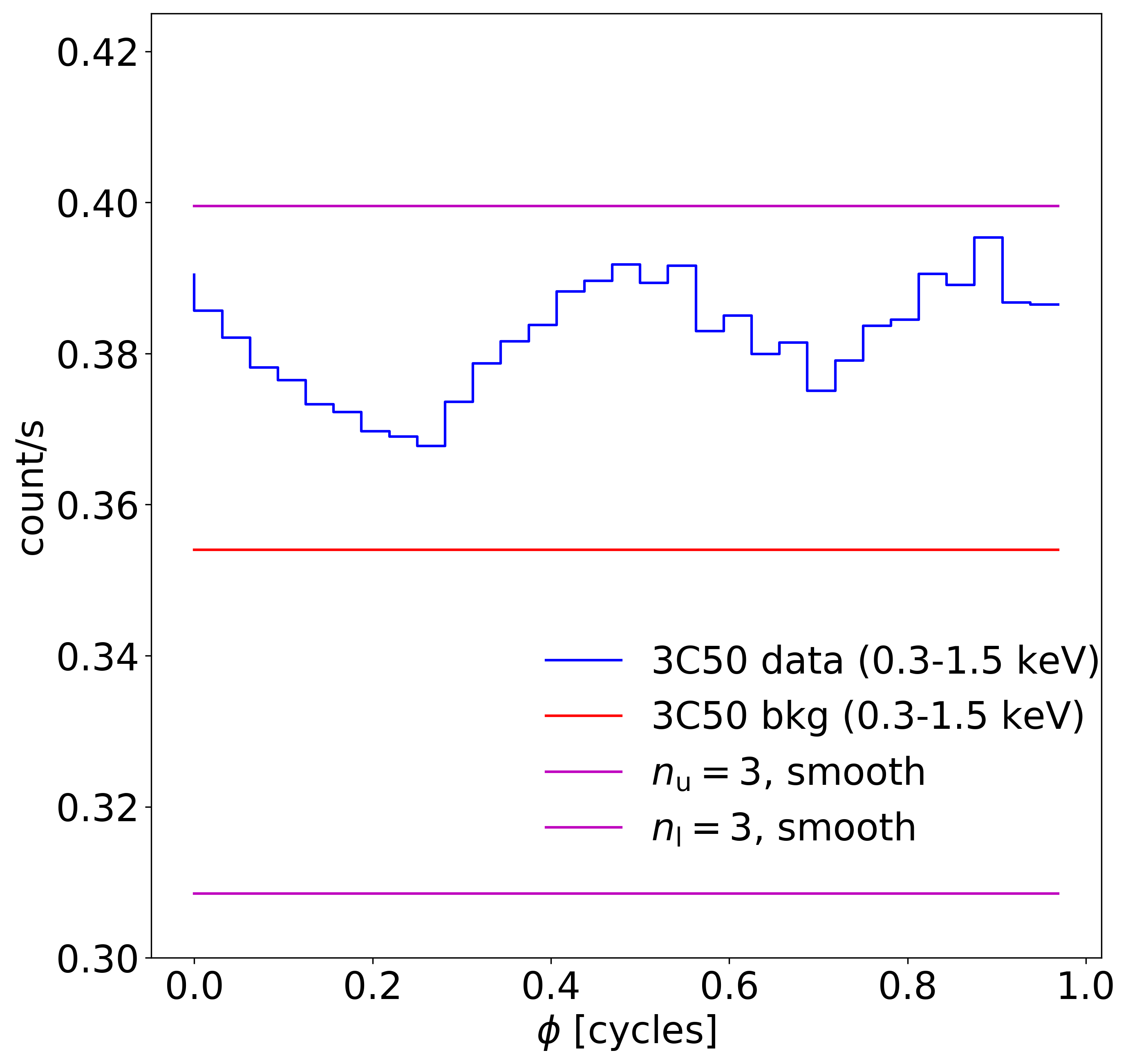}}
    \caption{\small{
    Prior limits for the background of the 3C50 data set for the bolometric pulse profile. 
    The energy-integrated count-rate pulse profile and the 3C50 background estimate are shown with blue and red step functions. 
    The magenta step functions show the smoothed $3 \sigma$ lower and upper limits for the energy-integrated background, corresponding to those shown in Figure \ref{fig:3c50_bkgs}. 
    The complete figure set (4 images) for different energy intervals are shown in the online journal (HTML version). 
    }
    }
    \label{fig:3c50_bkgs_pulse}
    \end{figure}
}

The third (and new) case we consider is the analysis of \NICER data with tighter lower and upper limits, $\{\mathfrak{L}\}$ and $\{\mathfrak{U}\}$, for the background based on a \NICER background estimate. 
For the fourth case (with \xmm observations included as in Equation \eqref{eqn:background-marginalized likelihood}), we describe the likelihood as
\begin{align}
\begin{split}
    & p(d_{\rm N}, d_{\rm X}, \{\mathcal{B}_{\rm N}\}, \{\mathscr{B}_{\rm X}\} \,|\, s)
    \propto
    \mathop{\int}_{\{\mathfrak{L}\}}^{\{\mathfrak{U}\}}
    p(d_{\rm N} \,|\, s, \{\mathbb{E}[b_{\rm N}]\}, \textsc{nicer}) \\
    & \times
    d\{\mathbb{E}[\mathcal{B}_{\rm N}]\}
    \mathop{\int}_{\{\mathscr{L}\}}^{\{\mathscr{U}\}}
    p(d_{\rm X} \,|\, s, \{\mathbb{E}[b_{\rm X}]\}, \textsc{xmm})d\{\mathbb{E}[\mathscr{B}_{\rm X}]\},
    \label{eqn:background-marginalized_likelihood_new}
\end{split}
\end{align}
where $\{\mathcal{B}_{\rm N}\}$ is the set of \NICER background count rates from the background estimate.\footnote{
The third case corresponds to a similar equation as \eqref{eqn:background-marginalized_likelihood_new} but without dependence and integration over \xmm background variables. 
}
We note that $\{\mathcal{B}_{\rm N}\}$ is already defined in count rates directly comparable to the \NICER data, and does not need to be rescaled using BACK\_SCAL or exposure time factors, unlike $\{\mathscr{B}_{\rm X}\}$.

The integration limits $\{\mathfrak{L}\}$ and $\{\mathfrak{U}\}$ were determined either based on the SW estimate (only lower limit) or on the 3C50 background estimate (either only lower or both lower and upper limits); see Section \ref{sec:model_sw} and Appendix \ref{sec:3C50_data_and_bkg} for information about those models. 
In the case where we used only a lower limit from the \NICER background estimates, the upper limits (for each channel) were set to the same unspecified values $\{\mathscr{U}\}$ as in Equation  \eqref{eqn:background-marginalized likelihood}. 
This is motivated by the existence of known sources in the FoV of NICER's observation of \joh, which are not accounted for in the background estimates \citep{Wolff20}. 
However, we also explored the sensitivity of the results to a conservative upper limit in some of our models. 

In case of the SW estimate, we considered two different lower limits: one with a factor of 0.9 times the original SW estimate, and the other 0.9 times a smoothed version of the SW estimate. 
The latter was employed after determining that the original lower limit was forcing the total inferred \NICER background to be higher by limiting the background only at the channel 87 (corresponding to nominal energy 0.87 keV), where the background estimate peaks up while the \NICER data do not have a similar feature (see Section \ref{sec:nicer_SW} for the inferences). 
The smoothed and non-smoothed backgrounds are shown in Figure \ref{fig:sw}.
The smoothing of the background estimate was performed using a Savitzky–Golay filter with a window length of 17 (number of coefficients) and a fifth order polynomial,\footnote{Coded in Python with spectrum$\_$smooth = scipy.signal.savgol$\_$filter(spectrum, 17, 5).} after restricting the estimate to the energy channels considered for the analysis (from channel 30 to 150). 
We note that the choice of the smoothing technique and the values for the filter are not based on detailed analysis, but chosen so that the end result approximates a spectrum rebinned to minimize oversampling (corresponding to averaging over three adjacent energy bins).

In case of the 3C50 background estimate (and 3C50 data), we considered several options for the lower and upper limits of the background variables. 
As explained in Appendix \ref{sec:3c50_uncertainties}, the 3C50 background spectrum is accompanied by an estimate of the systematic error that is rescaled based on the background spectrum itself.
For the total standard deviation $\sigma_{\mathcal{B}_{\rm N}}$ of the
background estimate at a given channel, we use the square root of the quadratic sum of the systematic and statistical errors. 
The latter come from the convolved statistical errors of the library spectra, and are actually tiny compared to the systematic uncertainty.  
Given this, we set the lower bound for the background prior to $\mathcal{B}_{\rm N}-n_{\mathrm{l}} \sigma_{\mathcal{B}_{\rm N}}$ and the upper bound to $\mathcal{B}_{\rm N}+n_{\mathrm{u}}\sigma_{\mathcal{B}_{\rm N}}$, where $n_{\mathrm{l}}$ and $n_{\mathrm{u}}$ are settings that control the degree of conservatism. 
We considered cases with ($n_{\mathrm{l}}=3$, $n_{\mathrm{u}}=\infty$), ($n_{\mathrm{l}}=3$, $n_{\mathrm{u}}=3$), and ($n_{\mathrm{l}}=2$, $n_{\mathrm{u}}=3$). 
Here $n_{\mathrm{u}}=\infty$ means that the default upper limit $\mathcal{U}$ is used instead, and even using $n_{\mathrm{u}}=3$ is a very conservative choice, because it cuts off only solutions with background almost as high as the data, as seen from Figure \ref{fig:3c50_bkgs}. 
The background and its estimated errors are also compared against the energy-integrated pulses for the 3C50 data set in Figure \ref{fig:3c50_bkgs_pulse} (and in the associated online figure set).

In all of the high-resolution runs for 3C50 presented here, we used a smoothed background estimate, using the same smoothing technique as for the SW estimate. 
According to our initial lower-resolution runs, smoothing was again required to mitigate a strong dependence on the influence of one or very few channels. 
We also checked (with low-resolution runs) that smoothing by averaging over three adjacent bins yields very similar results to the smoothing method used in this paper. 
However, as seen from Figure \ref{fig:3c50_bkgs} the background estimate may still overpredict the background at high-energy channels; therefore we also considered models where the data $d_{\mathrm{N}}$ replaces the background estimate at those channels: in other words, the lower limit was set to $\min (\mathcal{B}_{\rm N}-n_{\mathrm{l}} \sigma_{\mathcal{B}_{\rm N}},d_{\mathrm{N}}-n_{\mathrm{l}} \sigma_{\mathcal{B}_{\rm N}})$
\footnote{We note that with this definition of background prior the same data affect both the prior and the likelihood factor when calculating the background-marginalized likelihood.}. 
Hereafter, the models using the minimum function as the lower limit are labeled as \textit{mdb} (minimum of the data and the background). 

\subsection{Posterior Computation}

\begin{deluxetable}{ll}[b]
\tablecaption{Model Parameters}
\tablehead{\multicolumn{1}{l}{Parameter} & \multicolumn{1}{l}{Description}}
\startdata
$M$ $[\textit{M}_{\odot}]$ &
gravitational mass \\
$R_{\textrm{eq}}$ $[$km$]$ &
coordinate equatorial radius \\
$\Theta_{p}$ $[$radians$]$ &
$p^{\mathrm{a}}$ region center colatitude  \\
$\Theta_{s}$ $[$radians$]$ &
$s^{\mathrm{b}}$ region center colatitude \\
$\phi_{p}$ $[$cycles$]$ &
$p$ region initial phase$^{\mathrm{c}}$  \\
$\phi_{s}$ $[$cycles$]$ &
$s$ region initial phase$^{\mathrm{d}}$\\
$\zeta_{p}$ $[$radians$]$ &
$p$ region angular radius   \\
$\zeta_{s}$ $[$radians$]$ &
$s$ region angular radius \\
$\log_{10}\left(T_{p}\;[\textrm{K}]\right)$ &
$p$ region \TT{NSX} effective temperature \\
$\log_{10}\left(T_{s}\;[\textrm{K}]\right)$ &
$s$ region \TT{NSX} effective temperature \\
$\cos(i)$ &
cosine Earth inclination to spin axis \\
$D$ $[$kpc$]$ &
Earth distance \\
$N_{\textrm{H}}$ $[10^{20}$cm$^{-2}]$ &
interstellar neutral H column density \\
$\alpha_{\rm{NICER}}$ &
\NICER effective-area scaling \\
$\alpha_{\rm{XMM}}$ &
\xmm effective-area scaling \\
\enddata
\tablecomments{\ \\  $^{\mathrm{a}}$ Primary. \\ $^{\mathrm{b}}$ Secondary. \\ $^{\mathrm{c}}$ With respect to the meridian on which Earth lies. \\ $^{\mathrm{d}}$ With respect to the meridian on which the Earth antipode lies.}
\end{deluxetable}\label{table:all_params}

We use \XPSI to calculate the likelihood function and prior PDFs and then employ nested sampling to compute the posterior samples using \MultiNest \citep{MultiNest_2008,MultiNest_2009, PyMultiNest}. 
The details of the sampling and the resolution settings follow accurately the production calculations of \citetalias{Riley2021}: $4\times10^{4}$ live points; a bounding hypervolume expansion factor of $0.1^{-1}$; and an estimated remaining log-evidence of $10^{-1}$. 
For most of the models, we also performed first a low-resolution run with $4\times10^{3}$ live points, where the inferred median NS radius was typically 0.2-0.4 km smaller, and the credible interval was more constrained than in the high-resolution runs. 
This was also noted in \citetalias{Riley2021}, where a small further broadening of the radius posterior was seen even if using $8\times10^{4}$ live points. 
Due to the additional computation expense and the aim of studying mainly the sensitivity of the results to different background modeling choices (which appear more significant than the sensitivity to increasing resolution further), we still consider $4\times10^{4}$ live points as a suitable resolution for the main runs of this paper. 
This resolution is used in all the runs unless stated otherwise (a couple of sensitivity tests referred as \textit{low-resolution} use $4\times10^{3}$ live points instead). 
All our free sampling parameters are shown and described in Table \ref{table:all_params}. 

\section{Inferences}\label{sec:results}

In this section, we report our results. 
First we show the effects of using the SW background estimate in the \NICER analysis without applying \xmm data (case (3) from Table \ref{table:cases}). 
Then we present our inference based on the 3C50 filtering of the \NICER data, first without using the corresponding background estimate (case (1) from the same table) and then including the estimate using several different methods (case (3)). 
We also perform a few joint 3C50 and \xmm runs (applying both cases (2) and (4)). 
We finish the section by showing the results of runs with antipodal models and measuring the degree of non-antipodality of the other runs. 

\subsection{Effect of the SW Estimate }\label{sec:nicer_SW}

{
\begin{figure}[t!]
\centering
\resizebox{\hsize}{!}{\includegraphics[
width=\textwidth]{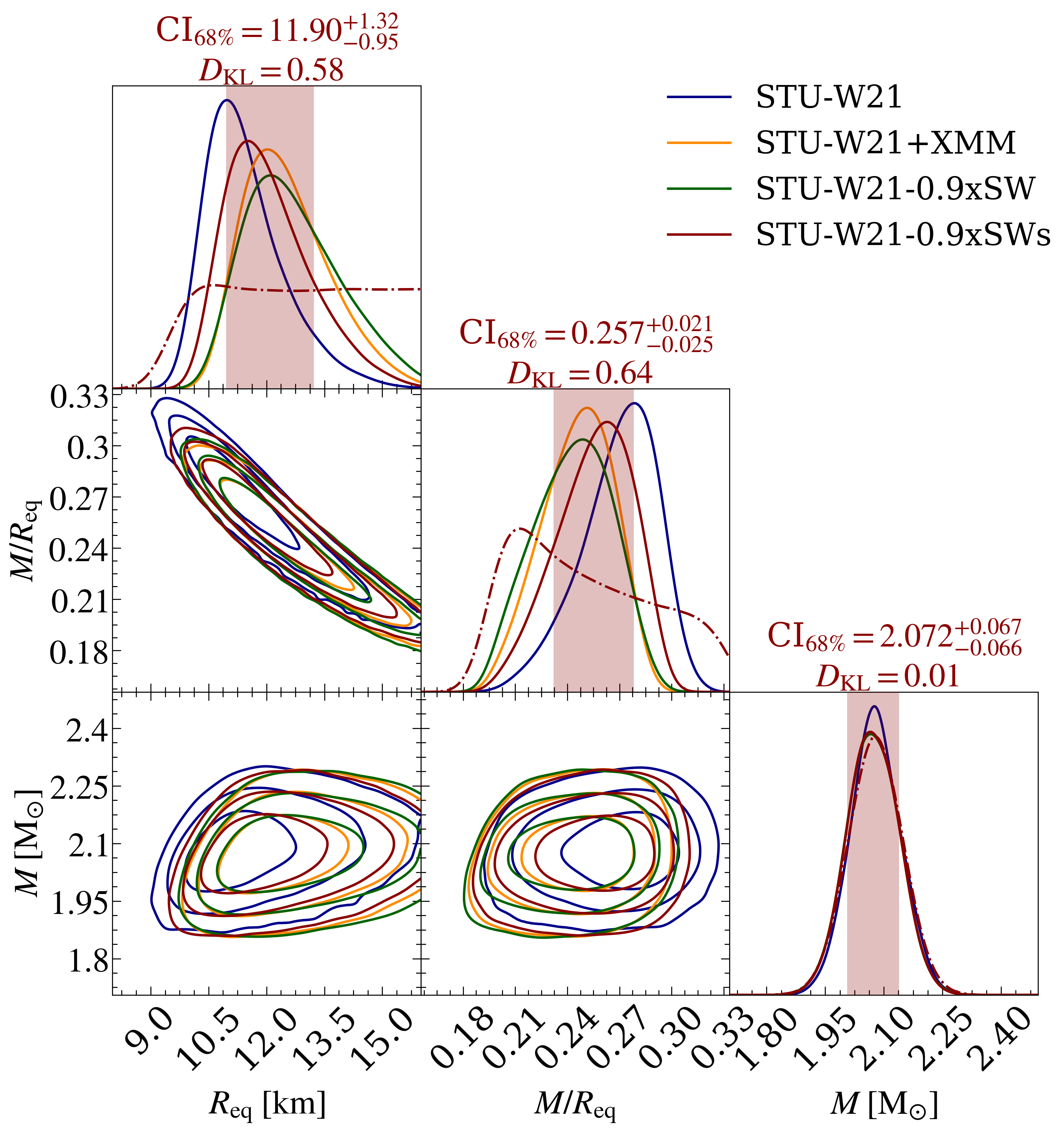}}
\caption{\small{
    Effect of the SW background estimate on the mass and radius posterior distributions using the W21 data set conditional on the \texttt{ST-U} model.
    Also, the posteriors for compactness $GM/R_{\rm eq}c^{2}$ are shown (hereafter and in the figure referred as $M/R_{\rm eq}$, i.e., assuming $c=1$ and $G=1$). 
    Four types of posterior distribution are shown: one conditional on the \NICER likelihood function with a smoothed SW background lower bound (STU-W21-0.9xSWs); one similar but with a non-smoothed lower bound (STU-W21-0.9xSW); one conditional on \NICER likelihood function without background constraints (STU-W21); and one conditional on the \NICER and \xmm likelihood function (STU-W21+XMM). 
    The last two are from \citetalias{Riley2021}.
    The marginal prior PDFs are shown by the dashed-dotted functions. 
    We report the credible intervals and the Kullback–Leibler divergence $D_{\mathrm{KL}}$ for the \NICER posterior conditional on the smoothed SW estimate lower bound (for the other cases, see Table \ref{table:sw}).  
    The shaded intervals contain $68.3\%$ of the posterior mass, and the contours in the off-diagonal panels contain $68.3\%$, $95.4\%$, and $99.7\%$ of the posterior mass. 
    See the caption of Figure 5 of \citetalias{Riley2021} for additional details about the figure elements. 
}}
\label{fig:spacetime_SW}
\end{figure}
}

{
    \begin{figure*}[t!]
    \centering
    {\includegraphics[
    width=0.49\textwidth]{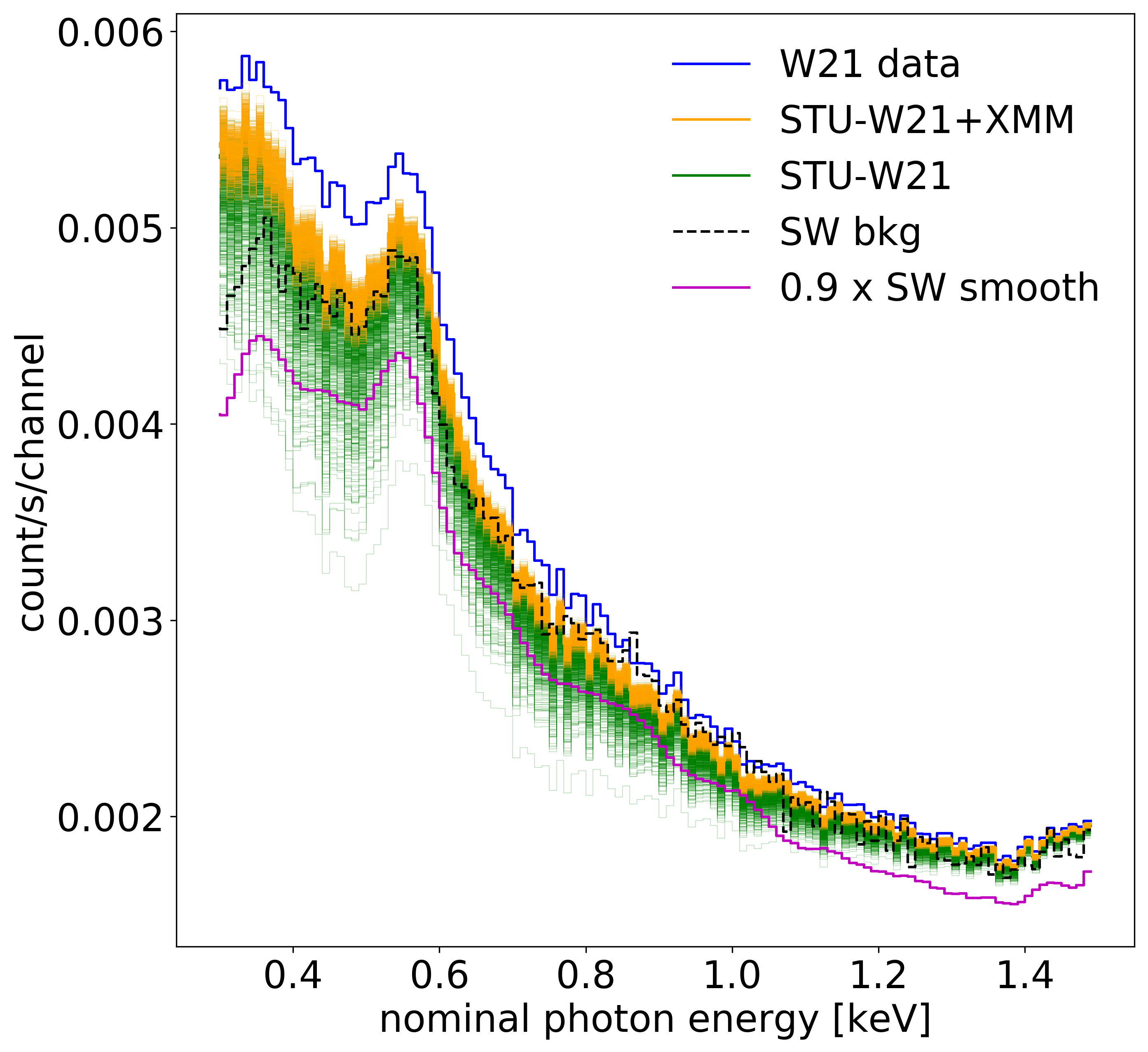}}
    {\includegraphics[
    width=0.49\textwidth]{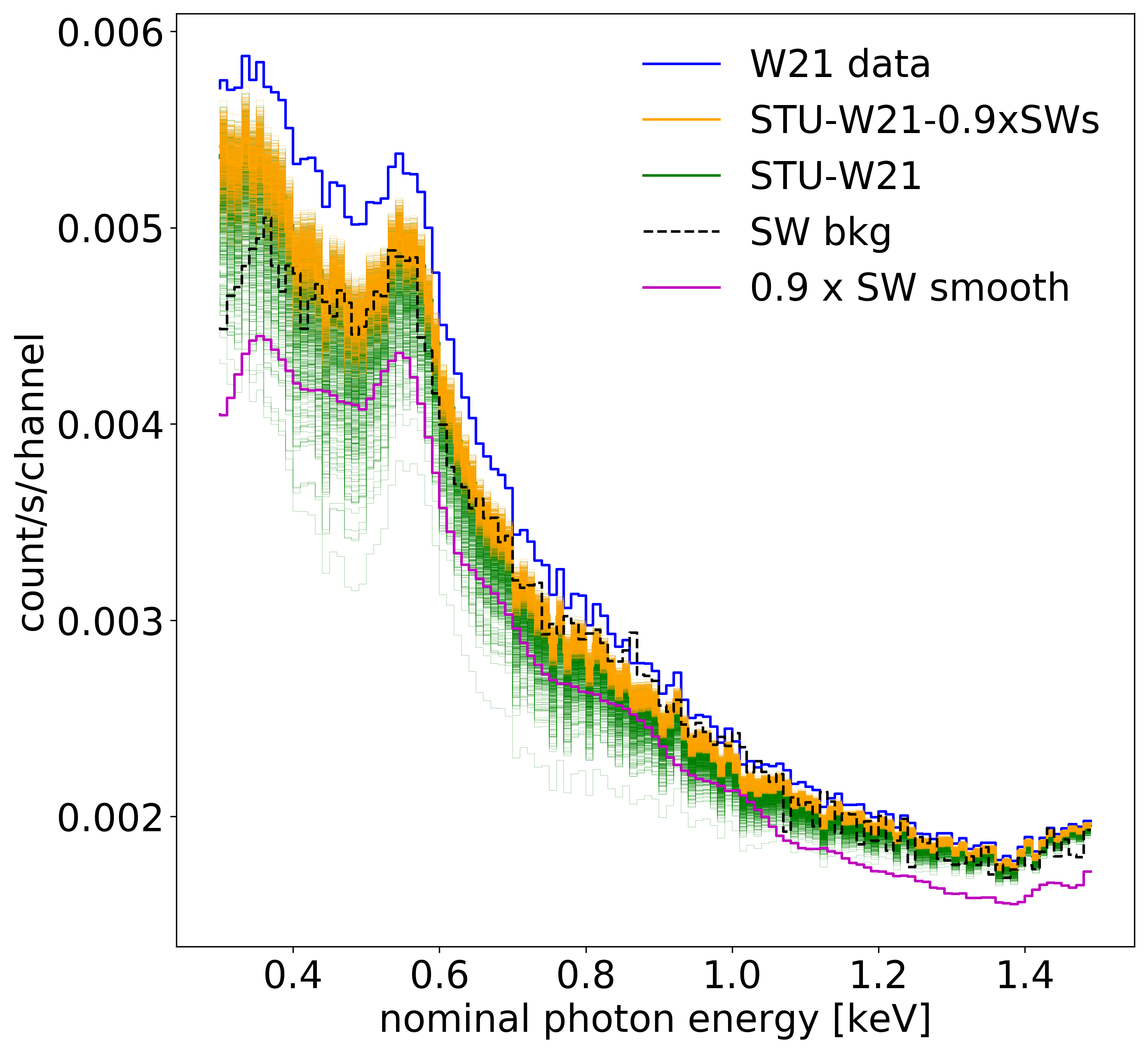}}
    \caption{\small{
    Comparison of the inferred \NICER background for the W21 data set based on different models. 
    Left panel: 
    the blue step function is the total \NICER count-rate spectrum, 
    the dashed black step function is the SW background estimate, and the magenta step function is the SW estimate multiplied by a factor of 0.9 and smoothed.
    Orange step functions show 1000 background curves that maximize the likelihood for 1000 equally weighted posterior samples from the previously published joint \NICER and \xmm run, as in Figure 15 of \citetalias{Riley2021}. 
    The green step functions show the corresponding inferred backgrounds for the previous \NICER-only analysis. 
    The orange functions are partly overlapping the green functions. 
    Right panel: 
    same as left panel, except the orange step functions present now the inferred background from the new \NICER-only run with the smoothed SW constraint. 
    The complete figure set (6 images), including also the results for the non-smoothed SW constraints, and inferred backgrounds based on the 1000 highest-likelihood samples, is available in the online journal (HTML version). 
    }}
    \label{fig:sw_bkg_inferred}
    \end{figure*}
}

As explained in Appendix \ref{sec:3C50_data_and_bkg} and Section \ref{sec:likelihood_background}, we studied the effect of using the SW estimate as a lower limit for the \NICER background, instead of applying a joint \NICER and \xmm analysis as in \citetalias{Riley2021}, \citet{Miller2021}.
The resulting posterior probability distributions for the spacetime parameters (mass, radius, compactness) are shown in Figure \ref{fig:spacetime_SW}, using either smoothed or non-smoothed 0.9 times the SW estimate as the lower limit (a summary can be found in Table \ref{table:sw} and Figure \ref{fig:radius_intervals} in Appendix \ref{apndx:tables}). 
Posteriors are compared against the headline results from \citetalias{Riley2021} which used the joint \NICER and \xmm analysis with non-compressed effective-area scaling between the instruments (and the incorrect BACK\_SCAL factor), giving a median radius of around 12.4 km (for the compressed scaling, it was 12.7 km). 
We see that the new \NICER-only results, using the SW background information, are consistent with the previous joint \NICER and \xmm results, but the median radius depends slightly on whether we use the non-smoothed or smoothed SW estimate (see Section \ref{sec:likelihood_background} for definitions).
The former gives 12.5 km and the latter 11.9 km. 
We interpret this as indicating that \xmm is posing a stricter condition on the background than the 0.9 times the SW lower limit, rather than indicating the non-smoothed version being more accurate. 
However, both are higher than the median radius from the previous \NICER-only run, which was about 11.3 km.\footnote{The 11.3 km radius was obtained by applying importance sampling to include updated mass, inclination, and distance priors and presented in the Appendix of \citetalias{Riley2021}.} 

The increase in the radius, when constraining the background by setting a lower limit, can be understood from the necessity to decrease the unpulsed component emitted from the NS surface, in order to associate a larger fraction of the unpulsed emission with the background. An obvious way to decrease the unpulsed component is to decrease the NS compactness, so that fewer photons can reach the observer by light bending when the radiating spots are seen at high angles. 
Since the mass of \joh is tightly constrained by the prior, increasing the radius is the only way to decrease the compactness. 
As expected, the inferred background with the SW limit applied is consistent with the inferred \NICER background from the old joint \NICER and \xmm results.
This is seen in Figure \ref{fig:sw_bkg_inferred} by comparing the orange lines from the left and right panels, showing the background-constrained results.
The green lines show the original \NICER-only results, which are the same in both panels.
In both cases, the unpulsed NS component is decreased relative to the results inferred if no background constraints are applied. 
Smoothing the SW estimate affects the constraints, as it eliminates a few noisy peaks present in the SW estimate (e.g., at channel 87), that would otherwise force the entire inferred background to increase. 

By inspecting the posterior distributions of the other model parameters, we see only small changes in most of them (the full set of posterior figures for hot region parameters is provided in Section \ref{sec:nicer_3C50} of the online journal). 
This implies that the change in the background can be mainly compensated by adjusting the compactness. 
The inferred configuration (hot spots located near the equator) remains also very similar to that found in \citetalias{Riley2021}. 
However, we note that constraints especially on the spot colatitudes are not very stringent. 
That can also be noticed from the animations showing the maximum likelihood and maximum posterior configurations for different runs, presented later in Section \ref{sec:nicer_STS}. 

Furthermore, no significant difference in the model performance can be detected between the different background treatments. 
The residuals between the data and best-fit models can be found as part of the online journal figure set in Section \ref{sec:nicer_STS}. 
Posterior distributions for the other parameters (distance, cosine of observer inclination, effective-area scaling factor, and the hydrogen column density) are also included in the figure sets in Section \ref{sec:nicer_3C50}, and the expected pulse signals for each of the models can be found in the repository of \citet{salmi_zenodo22}. 

In addition, we explored the sensitivity to the choice of the factor 0.9 with low-resolution inference runs (using the smoothed version of the estimate), and found significant differences if using either 0.85 or 0.95 factor for the lower limit instead.
Using 0.85 times SW as the lower limit yields similar results as without any background constraints, and using 0.95 times SW shifts the median radius up to around 13.5 km. 
However, we think the latter limit potentially biases the results if the SW estimate were overpredicting the background strongly enough at even a few of the considered energy channels. 
We examine the sensitivity to different cutoffs in the background prior more thoroughly for the 3C50 data in Section \ref{sec:nicer_3C50_bkg_3C50}.

\subsection{Effect of Using 3C50 Data }\label{sec:nicer_3C50}

As explained in Appendix \ref{sec:3C50_data_and_bkg} and Section \ref{sec:likelihood_background}, we have analyzed another data set, called 3C50, produced using new filtering and providing improved background constraints. Since the data set itself is different, we first show the parameter constraints without applying the corresponding background estimates (case (1) in Table \ref{table:cases}), in the same manner as for the \NICER-only analysis in \citetalias{Riley2021} for the W21 data set.
The effect on the spacetime parameters, when using 3C50 filtered data (without background information) instead of the W21 data, is shown in Figure \ref{fig:spacetime_3C50} (comparing again to the same already published results as in Section \ref{sec:nicer_SW} and in Figure \ref{fig:spacetime_SW}), and the parameter values for the new run are given in Table \ref{table:3c50}. 
Using the 3C50 data set shifts the posterior distribution on the radius to be more consistent with that inferred previously from the joint \NICER and \xmm analysis. 
The uncertainty is however larger, which is partly expected due to the smaller number of detected counts for the 3C50 data set, which is 594709 instead of 628280 counts in the same energy intervals \citep[see][for discussions of the expected uncertainty as a function of counts and pulse amplitudes]{lomiller13,Psaltis2014}. 
However, that difference cannot fully explain the change, as discussed further in Section \ref{sec:radius_uncertainties}.

{
    \begin{figure}[t!]
    \centering
    \resizebox{\hsize}{!}{\includegraphics[
    width=\textwidth]{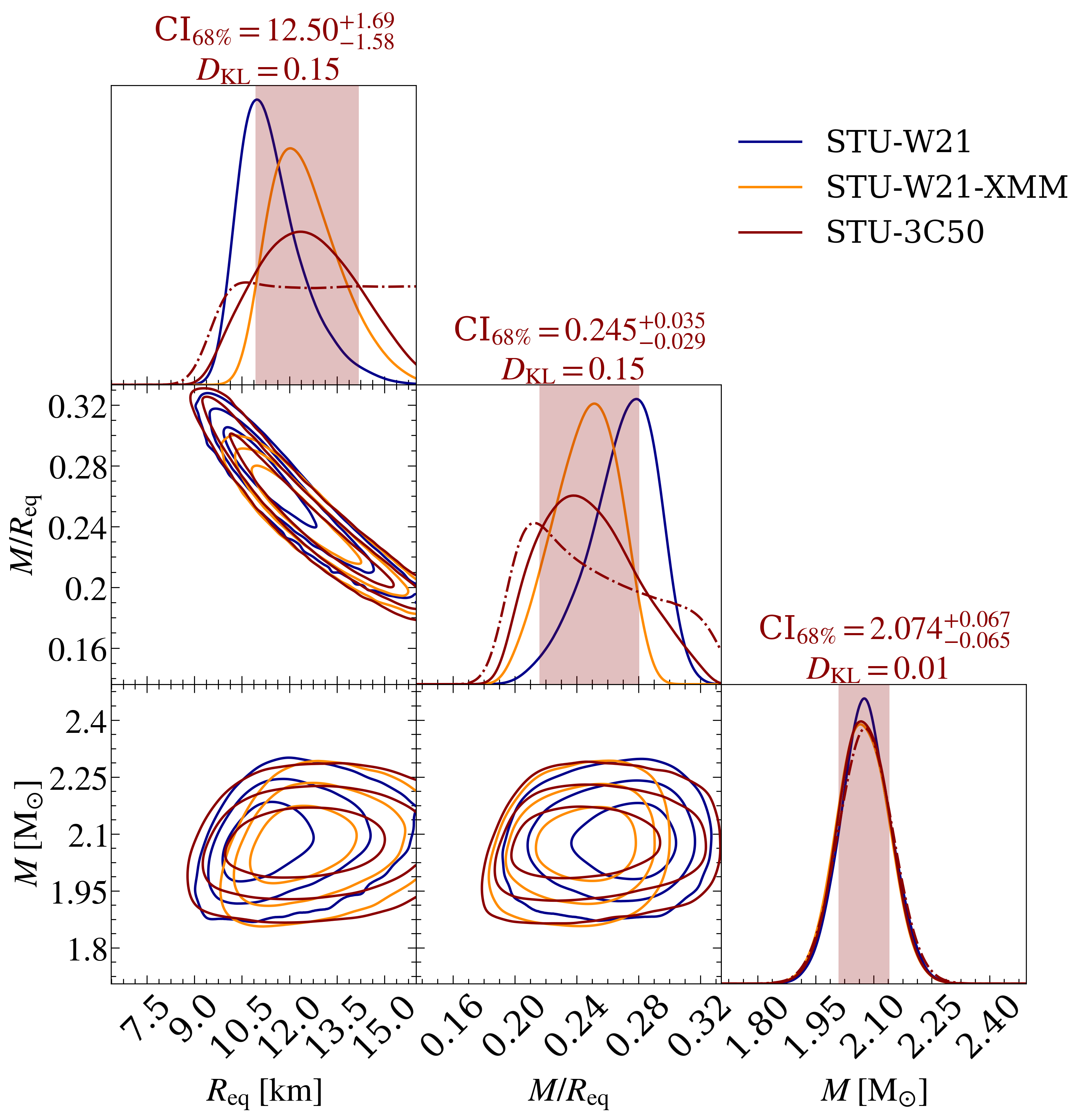}}
    \caption{\small{
    Effect of 3C50 data selection on the mass and radius posterior distributions conditional on the \texttt{ST-U} model. 
    Three types of posterior distribution are shown: one conditional on the \NICER likelihood function using 3C50 data set; one conditional on the \NICER likelihood function using the W21 data set; one conditional on \NICER W21 data and \xmm likelihood function (the latter two are the same as in Figure \ref{fig:spacetime_SW}). 
    We report the credible intervals and the divergence estimates for the \NICER posterior conditional on the 3C50 data set. 
    See the caption of Figure \ref{fig:spacetime_SW} for additional details about the figure elements. 
    }}
    \label{fig:spacetime_3C50}
    \end{figure}
}

The shift of the radius to higher values could be related to the 3C50 data set being cleaner than the W21 data set, and thus genuinely providing a radius estimate closer to the original joint W21 and \xmm result where the background is more constrained. 
However, we note that the shift is still relatively small compared to the statistical uncertainties in the measured radius. 
Similarly, a shift to a higher radius with 3C50 data set was also found in the independent analysis presented in Appendix \ref{apndx:A}, using the same PPM code and analysis pipeline as in \citealt{Miller2021}. 

The parameter constraints in the new 3C50 results are mostly similar to the original joint constraints, as seen from the posterior distributions of the hot region parameters in Figure \ref{fig:geometry_posteriors_all}.  
A notable exception is the relative phase shift produced by the different data filtering procedures.  
The hot spots are also slightly closer to the equator for the 3C50-only and W21-\xmm runs than for the original W21-only run. 
However, the 3C50 results differ from both previous results in the sizes and temperatures of the hot spots, predicting slightly colder and larger hot spots. 
The inferred hydrogen column density $N_{\mathrm{H}}$ for the 3C50 filtered data is significantly higher than that in the original results, which is shown in the posterior distributions of other model parameters in Figure \ref{fig:other_posteriors_all}. 
This may be explained by the differences in the data sets at the lowest energy channels, where the 3C50 selection has a lower count rate (as seen from Figures \ref{fig:sw} and \ref{fig:3c50_bkgs}) allowing stronger interstellar absorption.
The correlation between $N_{\mathrm{H}}$ and $R_{\mathrm{eq}}$ could also partly explain why the 3C50 filtering infers a slightly different NS radius (although seen only in the 2D posterior of $N_{\mathrm{H}}$ and $R_{\mathrm{eq}}$ for 3C50 case).
These parameters are known to be related due to their degeneracy with the temperature, at least in the case of spectral fitting as in \citet{GGS2019}.

{
    \begin{figure*}[t!]
    \centering
    \resizebox{\hsize}{!}{\includegraphics[
    width=\textwidth]
    {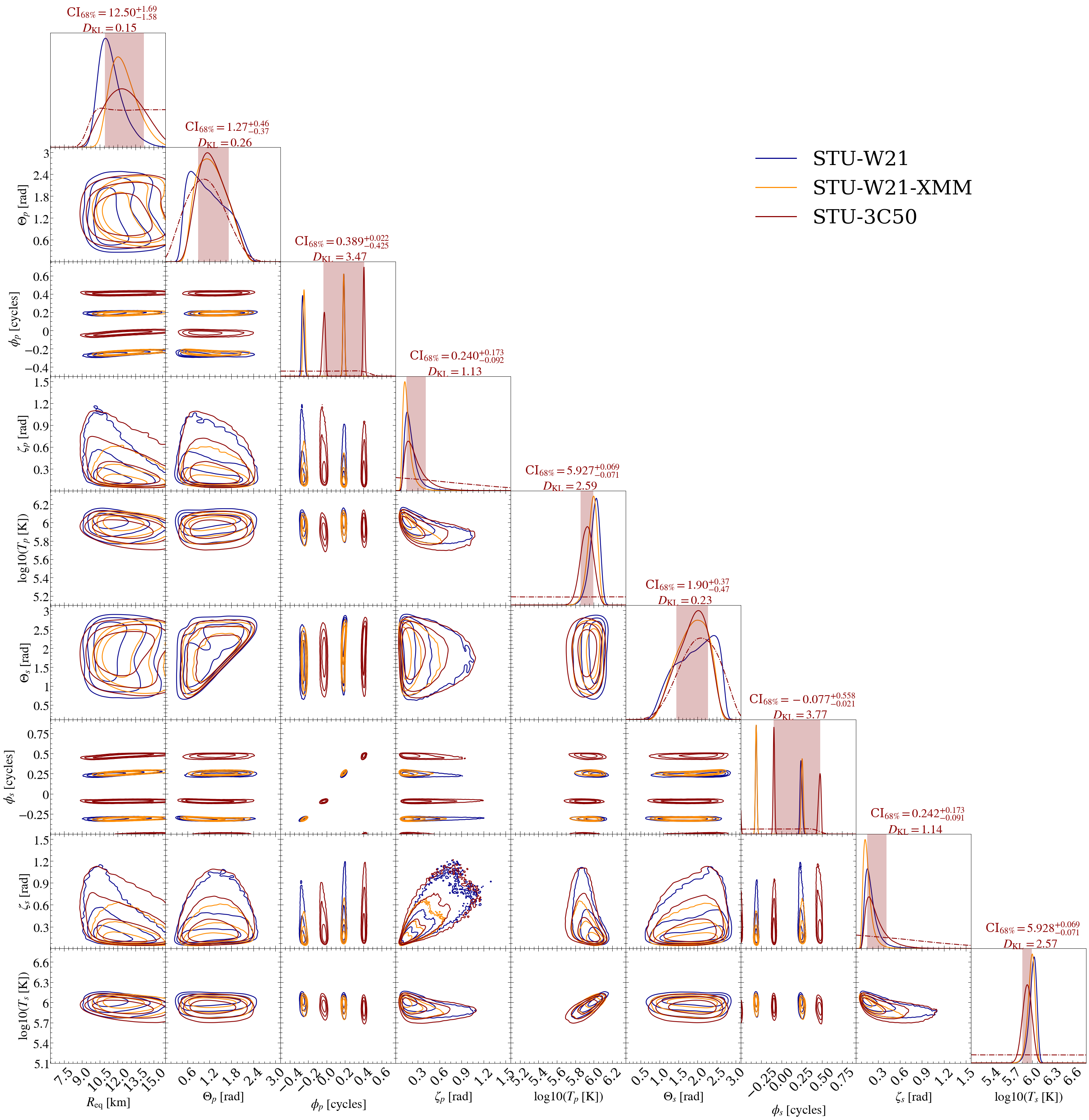}}
    \caption{\small{
    Effect of 3C50 data selection on the hot region parameter posterior distributions conditional on the \texttt{ST-U} model. 
    The same types of posterior distributions are shown as in Figure \ref{fig:spacetime_3C50}, and the shown parameters are described in Table \ref{table:all_params}. 
    The credible intervals and the divergence estimates are reported for the \NICER posterior conditional on the 3C50 data set. 
    The complete figure set (6 images), including also the results based on different background constraints and spot models, is available in the online journal (HTML version). 
    }}
    \label{fig:geometry_posteriors_all}
    \end{figure*}
}

{
    \begin{figure*}[t!]
    \centering
    \resizebox{\hsize}{!}{\includegraphics[
    width=\textwidth]
    {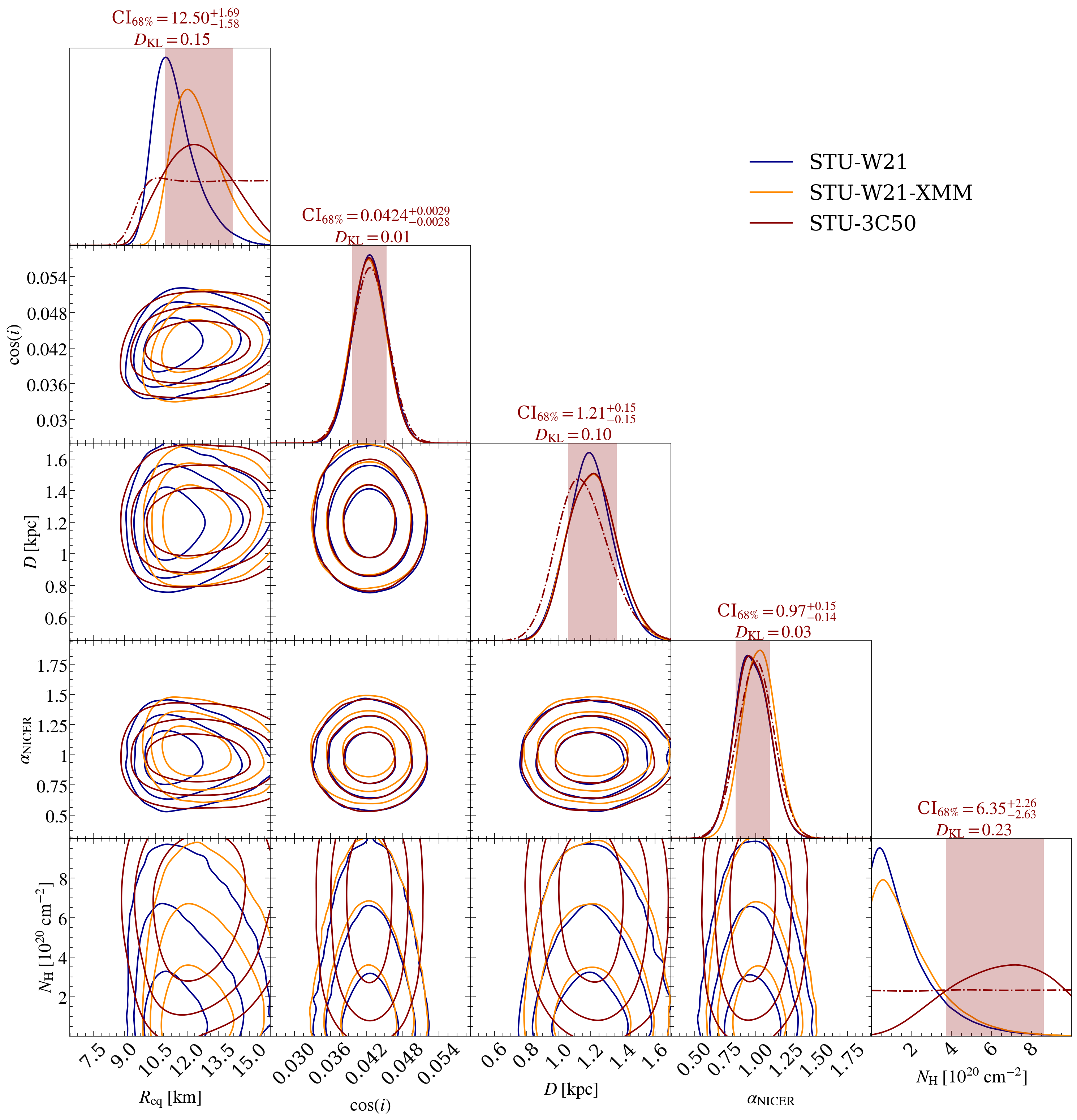}}
    \caption{\small{
    Effect of 3C50 data selection on the additional parameter posterior distributions conditional on the \texttt{ST-U} model. 
    The same types of posterior distributions are shown as in Figure \ref{fig:spacetime_3C50}, and the shown parameters are described in Table \ref{table:all_params}. 
    The credible intervals and the divergence estimates are reported for the \NICER posterior conditional on the 3C50 data set. 
    The complete figure set (6 images), including also the results based different background constraints and spot models, is available in the online journal (HTML version). 
    }}
    \label{fig:other_posteriors_all}
    \end{figure*}
}

\subsection{Effect of Using 3C50 Data with Background Constraints }\label{sec:nicer_3C50_bkg}

\subsubsection{3C50 Data with 3C50 Background Constraints}\label{sec:nicer_3C50_bkg_3C50}

In addition to the results shown in Section \ref{sec:nicer_3C50}, we have analyzed the 3C50-filtered \NICER data combined with the associated background constraints, as explained in Section \ref{sec:likelihood_background} (the case referred as (3) in Table \ref{table:cases}). 
The spacetime posterior distributions are shown in Figure \ref{fig:spacetime_3C50bkg} when using the smoothed 3C50 background limit.
In the corresponding online journal figure set similar posteriors are shown using the modified mdb background estimate mentioned in the end of Section \ref{sec:likelihood_background} (where the background estimate is limited to not exceed the data).

We see clear shifts in the distributions depending on the choices for the background limits; however all of the results are still statistically consistent (the median values and credible intervals for radius for all the runs are shown in Tables \ref{table:3c50} and \ref{table:3c50mdb}, and visualized in Figure \ref{fig:radius_intervals}).  





{
    \begin{figure}[t!]
    \centering
    \resizebox{\hsize}{!}{\includegraphics[
    width=\textwidth]{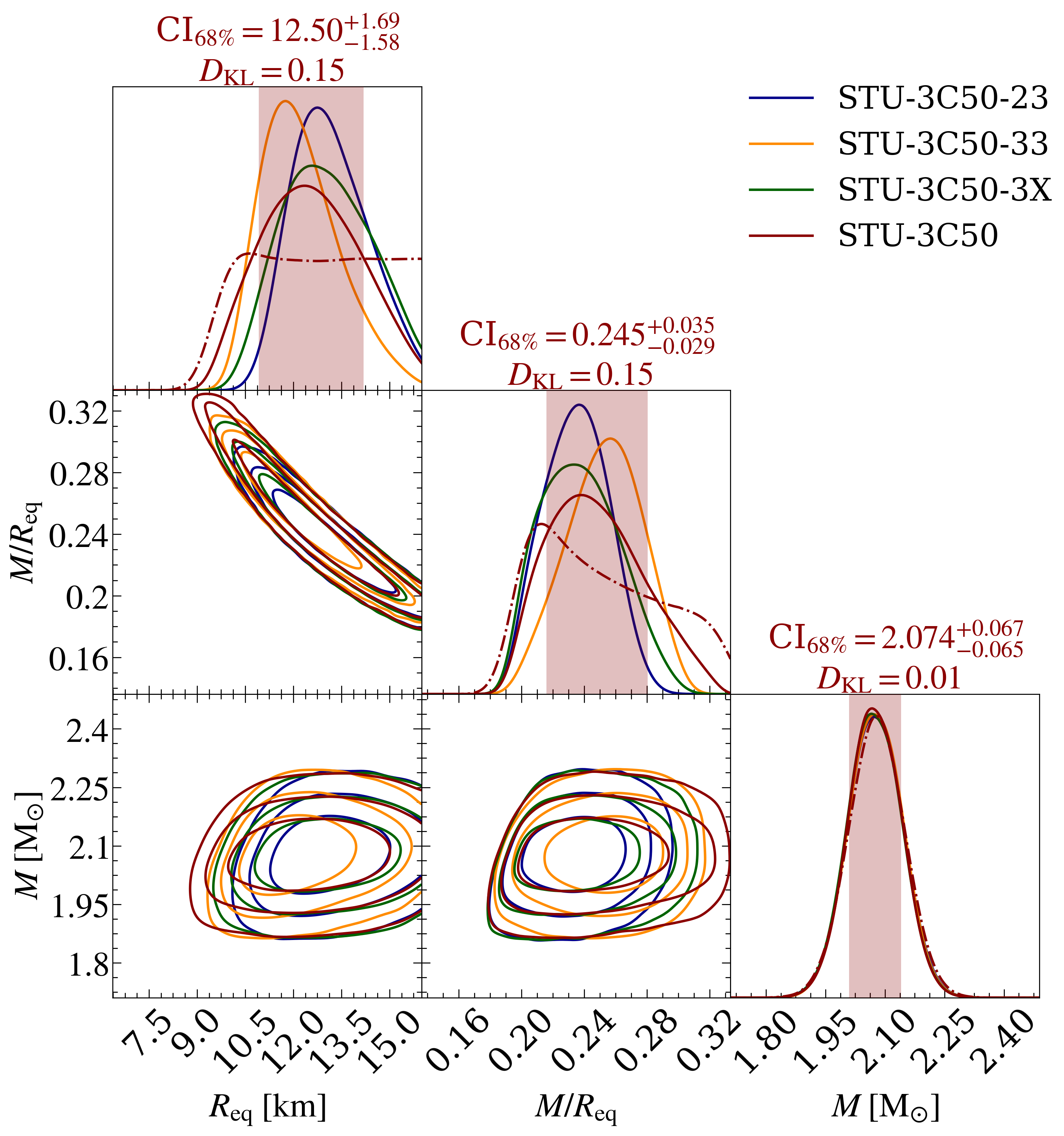}}
    \caption{\small{
    Effect of 3C50 background estimate on the mass and radius posterior distributions.
    Four types of  posterior distribution are shown: one conditional on the \NICER likelihood function using the \texttt{ST-U} model and 3C50 data set (same as in Figure \ref{fig:spacetime_3C50}); one otherwise same but with $n_{\mathrm{l}}=3$ lower limit for the background (3C50-3X); one otherwise same but with $n_{\mathrm{l}}=3$ lower and $n_{\mathrm{u}}=3$ upper limits for the background (3C50-33); and one otherwise same but with $n_{\mathrm{l}}=2$ lower and $n_{\mathrm{u}}=3$ upper limits for the background (3C50-23). 
    The credible intervals and the divergence estimates are reported for the \NICER posterior conditional on the 3C50 data set as in Figure \ref{fig:spacetime_3C50} (for the other cases, see Table \ref{table:3c50}). 
    See the caption of Figure \ref{fig:spacetime_SW} for additional details about the figure elements.  
    The corresponding posterior results using the lower limits based on the minimum of the data and the background (mdb) are shown in the online journal figure set (HTML version). 
    }}
    \label{fig:spacetime_3C50bkg}
    \end{figure}
}

From both posterior figures, we see that constraining the background using a lower limit always causes the radius to increase, and using an upper limit always causes it to decrease (compared to a case where the other limit has not been altered). 
For example, imposing $n_{l}=3$ lower limit shifts the median radius from 12.5 km to 13.0 km (in the case of Figure \ref{fig:spacetime_3C50bkg}). 
This effect is also similar to that found in the independent analysis of Appendix \ref{apndx:A}. 
Applying a $n_{u}=3$ upper limit in addition alters the median from 13.0 km to 12.0 km. 
In the case of a more restrictive lower limit, $n_{l}=2$, the median shifts from 12.0 km to 13.1 km. 
When using the mdb version of the estimate, the effects are similar but generally produce radii that are $0.3 - 0.7$ km smaller, due to the less strict lower limit of the background for a few noisy high-energy channels. 
In all cases, credible intervals of the radius tighten slightly when tighter background limits are applied. 
The difference between the $84 \%$ quantile and the median value shifts from about $\Delta R^{+}=1.6$ km to $\Delta R^{+}=1.4$ km, and the difference between the median and  $16 \%$ quantile shifts from about $\Delta R^{-}=1.7$ km to $\Delta R^{-}=1.1$ km, when considering the tightest background limits, compared to the results without any background constraints.  
Clearly, the credible intervals are most affected by the addition of the upper limit for the background.

The changes in the radius can be understood via compactness, as in Section \ref{sec:nicer_SW}: higher background leads to a higher radius (for a fixed mass) and vice versa. 
The inferred backgrounds are presented in Figure \ref{fig:3C50_bkg_inferred} for the case with $3 \sigma$ lower limit and no upper limit for the \NICER background. 
The other cases are shown in the corresponding online journal figure set. 
We see that setting either lower or upper limits to the background affects the inferred background (the orange step functions) by pushing it either up or down, as expected.
In addition, the inferred background (and the radius) is also moderately sensitive to changes in the prior limits in only a few relevant energy channels. This is demonstrated by the difference between the inferred backgrounds when using the smoothed mdb estimate versus the smoothed estimate without the minimum function (e.g., by comparing the Figures 1 and 4 in the online set).

{
    \begin{figure}[t!]
    \centering
    \resizebox{\hsize}{!}{\includegraphics[
    width=\textwidth]{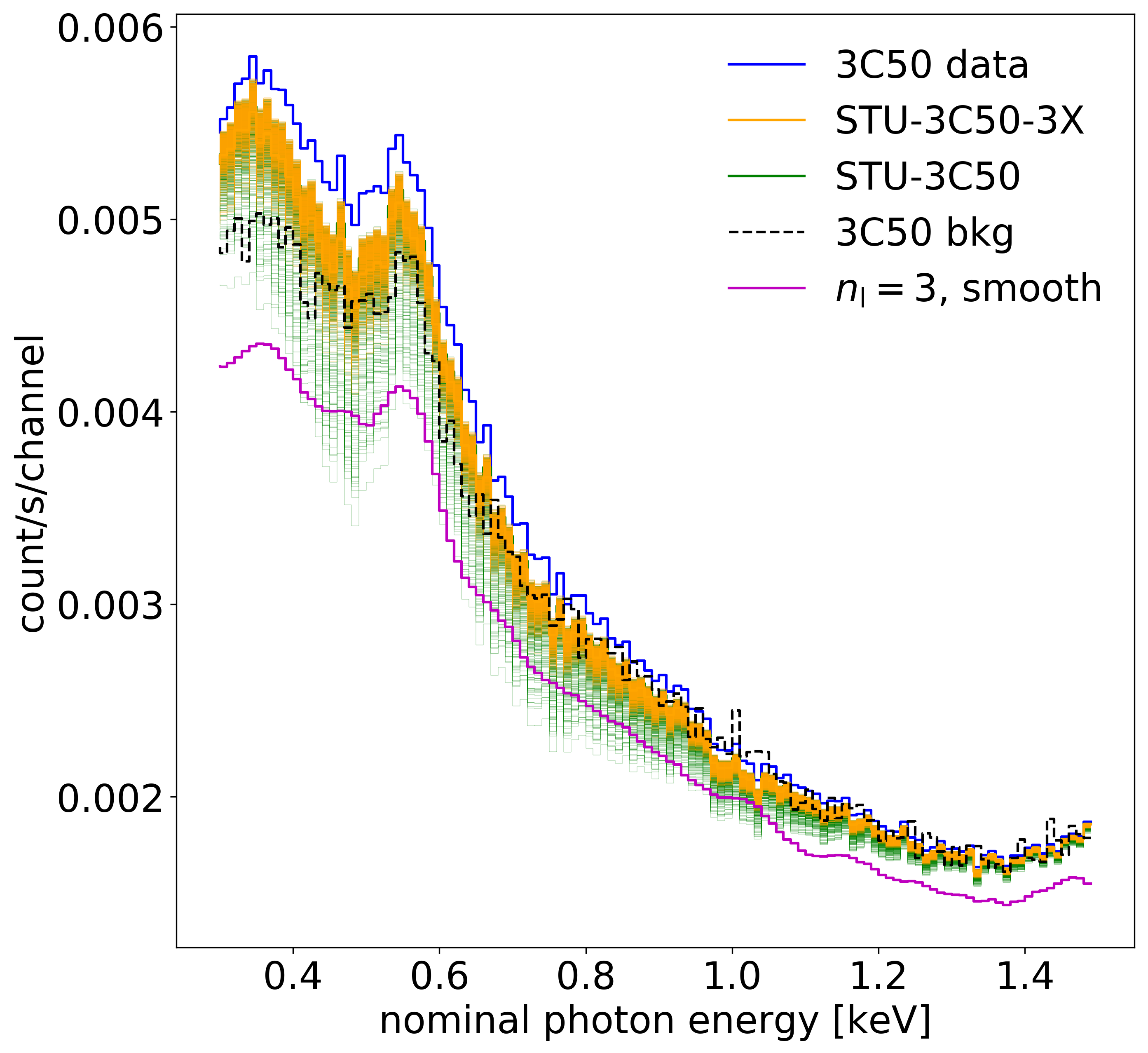}}
    \caption{\small{
Comparison of the inferred \NICER background for the 3C50 data set with and without background constraints.  
The blue step function is the total \NICER count-rate spectrum, 
the dashed black step function is the 3C50 background estimate, and the magenta step functions show the lower and upper bounds for background support. 
Orange step functions show 1000 background curves that maximize the likelihood for 1000 equally weighted posterior samples from the 3C50 data analysis when applying $n_{\mathrm{l}}=3$ lower limit for the \NICER background (with no upper limit).  
The green step functions show the corresponding inferred backgrounds for the 3C50 analysis without applying any background constraints.
The complete figure set (11 images) is available in the online journal (HTML version). These include the inferred backgrounds for the other background prior choices for 3C50. 
    }}
    \label{fig:3C50_bkg_inferred}
    \end{figure}
}

\subsubsection{3C50 Data and Background with \xmm} \label{sec:nicer_3C50_bkg_XMM}

We have also analyzed the 3C50-filtered \NICER data in combination with the same \xmm data set that was used in the previous analysis of \citetalias{Riley2021} (case (2) in Table \ref{table:cases}). In addition, we have tried constraining the \NICER background using both the \xmm data set and the 3C50 background estimate simultaneously (case (4)). 
The posterior distributions of spacetime parameters are shown in Figure \ref{fig:spacetime_3C50XMM} (and the credible intervals for each run are presented in Table \ref{table:3c50_XMM} and Figure \ref{fig:radius_intervals}).
From there we see that the inclusion of \xmm data shifts the median radius from 12.5 km to 12.9 km and narrows the width of the 68 \% credible interval of radius from 3.3 km to 2.2 km. 
Constraining the \NICER background further using the 3C50 estimate does not significantly affect the results, since the background is essentially more restricted using \xmm than providing, for example, background limits of $n_{\mathrm{l}}=2$ and $n_{\mathrm{u}}=3$ based on the 3C50 estimate (as seen from the full online set for Figure \ref{fig:3C50_bkg_inferred}). 
However, we note that these conclusions can depend on the assumed cross-calibration uncertainty between \NICER and \xmm, and for other sources or future data sets, the 3C50 uncertainties may be smaller. 
We use the compressed scaling for cross-calibration as explained in Section \ref{sec:instru_response}, which yielded an NS radius of $12.71^{+1.25}_{-0.96}$~km in \citetalias{Riley2021}, very similar to the 3C50+\xmm results obtained here (and also fairly close to the 3C50+\xmm results of Appendix \ref{apndx:A}).   
We also tested the sensitivity to the energy-independent cross-calibration uncertainty with low-resolution runs (as mentioned in Section \ref{sec:instru_response}), but found only relatively small effects on the inferred parameters (results shown in the online set of Figure \ref{fig:spacetime_3C50XMM}). 




{
    \begin{figure}[t!]
    \centering
    \resizebox{\hsize}{!}{\includegraphics[
    width=\textwidth]{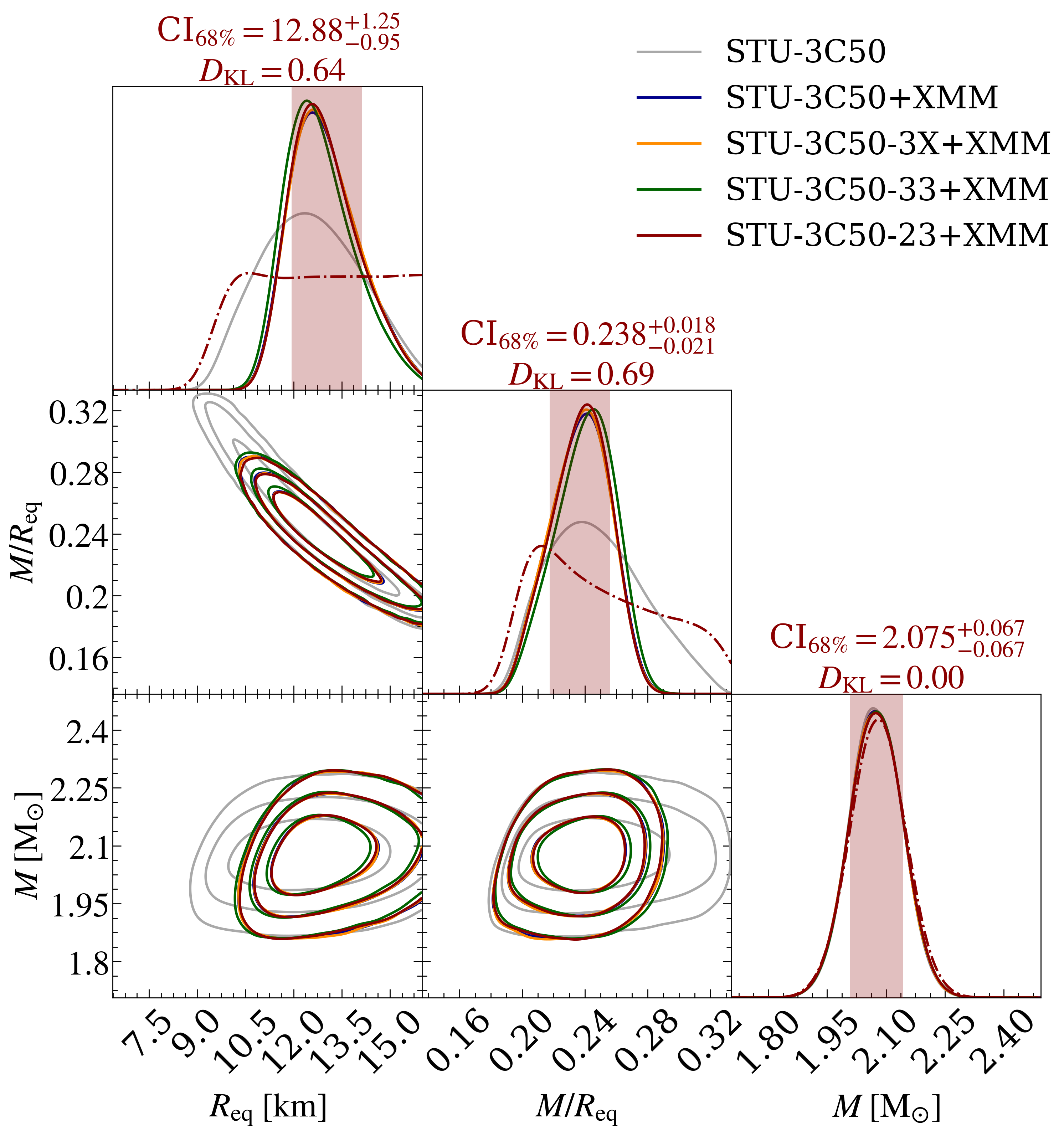}}
    \caption{\small{
    Effect of adding \xmm data to the 3C50 data on the mass and radius posterior distributions conditional on the \texttt{ST-U} model.
    Five types of posterior distribution are shown:
    one conditional on 3C50; one conditional on 3C50 and the \xmm likelihood function; one conditional on 3C50 with ($n_{\mathrm{l}}=3$ and $n_{\mathrm{u}}=\infty$) background limits and \xmm likelihood function; one conditional on 3C50 with ($n_{\mathrm{l}}=3$ and $n_{\mathrm{u}}=3$) background limits and the \xmm likelihood function; and one conditional on 3C50 with ($n_{\mathrm{l}}=2$ and $n_{\mathrm{u}}=3$) background limits and \xmm likelihood function.
    The credible intervals and the divergence estimates are reported for the \NICER posterior conditional on the 3C50 and the \xmm likelihood function (for the other cases, see Table \ref{table:3c50_XMM}). 
    See the caption of Figure \ref{fig:spacetime_SW} for additional details about the figure elements. 
    The complete figure set (2 images), which include also the results with different cross-calibration prior assumptions, is available in the online journal (HTML version). 
    }}
    \label{fig:spacetime_3C50XMM}
    \end{figure}
}

\subsection{Effect of Antipodality}\label{sec:nicer_STS}

As mentioned in Section \ref{sec:surf_hot_regions}, we have also analyzed models with less complexity in the surface hot regions, restricting the spots to be in an antipodal configuration. 
We applied both the models \texttt{ST-S} and \texttt{ST-Ua} (the former restricting also the temperatures and sizes of the spot to be the same), and compared against the non-antipodal \texttt{ST-U} results for the 3C50 data without background constraints (case (1)), and for the 3C50 data with ($n_{l}=3$, $n_{u}=\infty$) background limits combined with \xmm observation (case (4)). 
The resulting spacetime parameter constraints are shown in Table \ref{table:3c50_STS} and in Figure \ref{fig:spacetime_3C50_STS} for the background limited case. 
The residuals for the different cases are shown in Figure \ref{fig:resid0}.

{
    \begin{figure}[t!]
    \centering
    \resizebox{\hsize}{!}{\includegraphics[
    width=\textwidth]{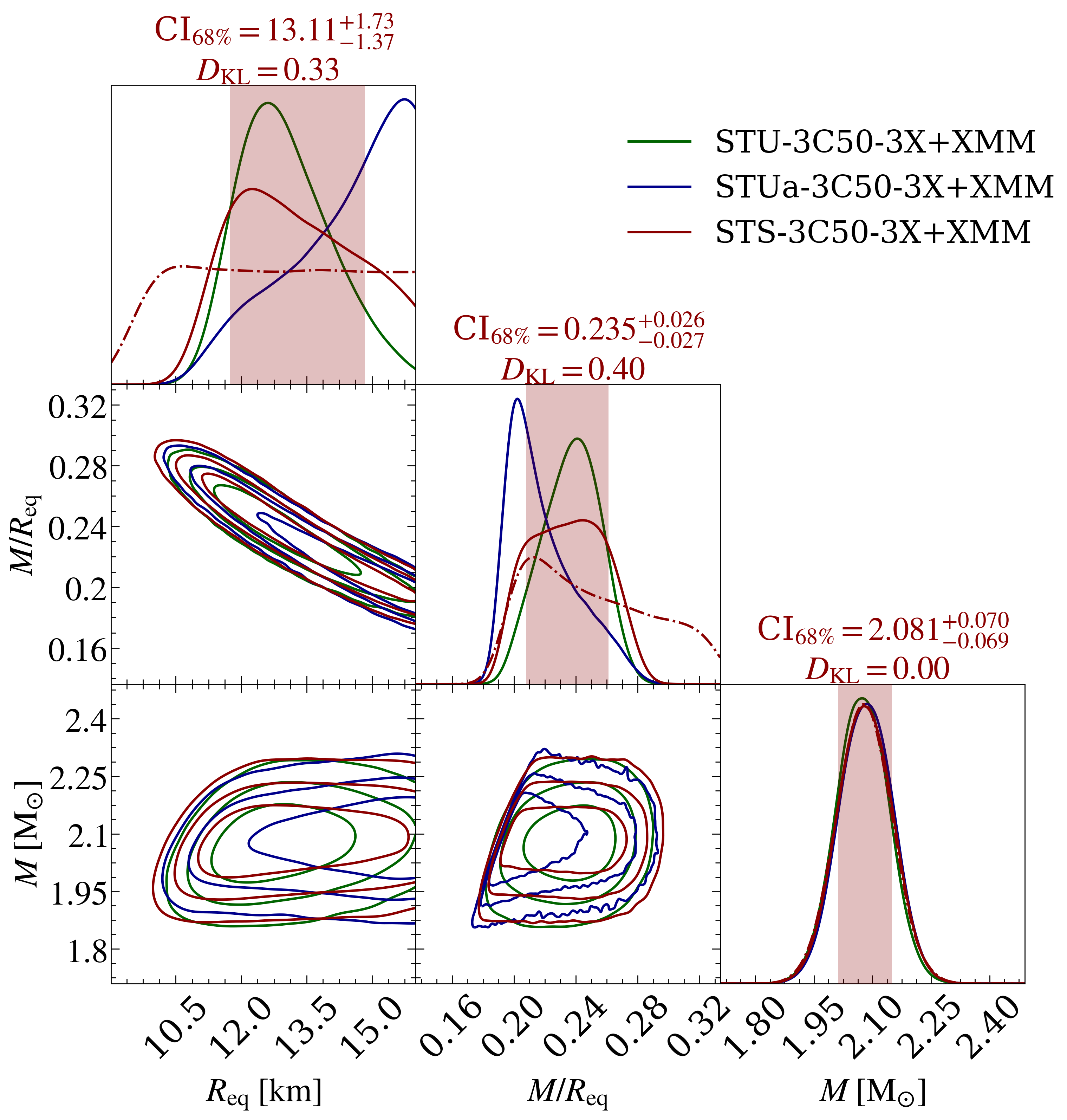}}
    \caption{\small{
    Effect of antipodality on the mass and radius posterior distributions conditional on 3C50 with ($n_{\mathrm{l}}=3$ and $n_{\mathrm{u}}=\infty$) background limits and \xmm likelihood function.
    Three types of posterior distribution are shown:
    one conditional on the \texttt{ST-U} model; one conditional on the \texttt{ST-Ua} model; and one conditional on the \texttt{ST-S} model. 
    The credible intervals and the divergence estimates are reported for the \NICER posterior conditional on the \texttt{ST-S} model (for the other cases, see Table \ref{table:3c50_STS}). 
    See the caption of Figure \ref{fig:spacetime_SW} for additional details about the figure elements.    
    }}
    \label{fig:spacetime_3C50_STS}
    \end{figure}
}






{
    \begin{figure*}[t!]
    \centering
    {\includegraphics[
    width=0.49\textwidth]{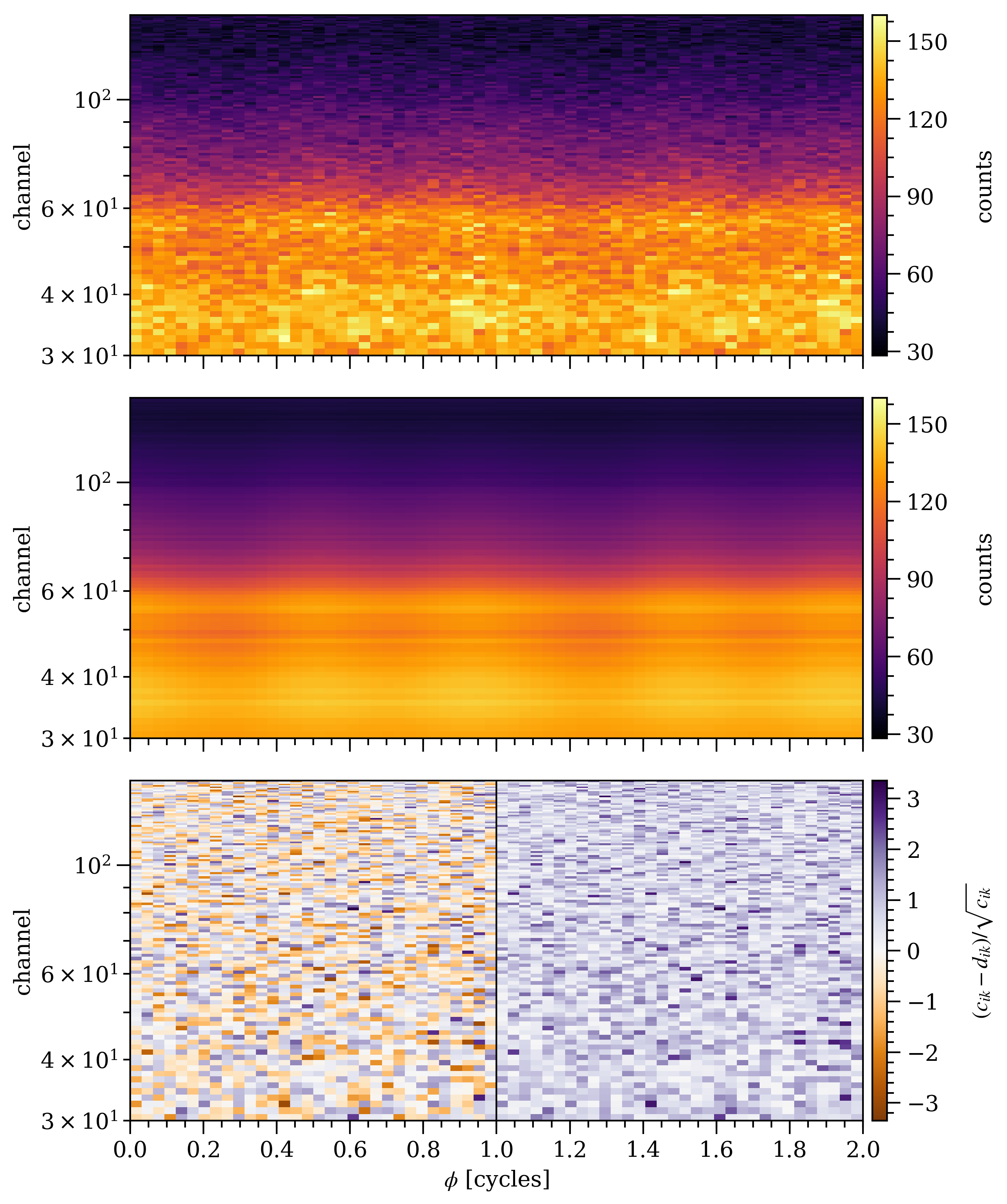}}
    {\includegraphics[
    width=0.49\textwidth]{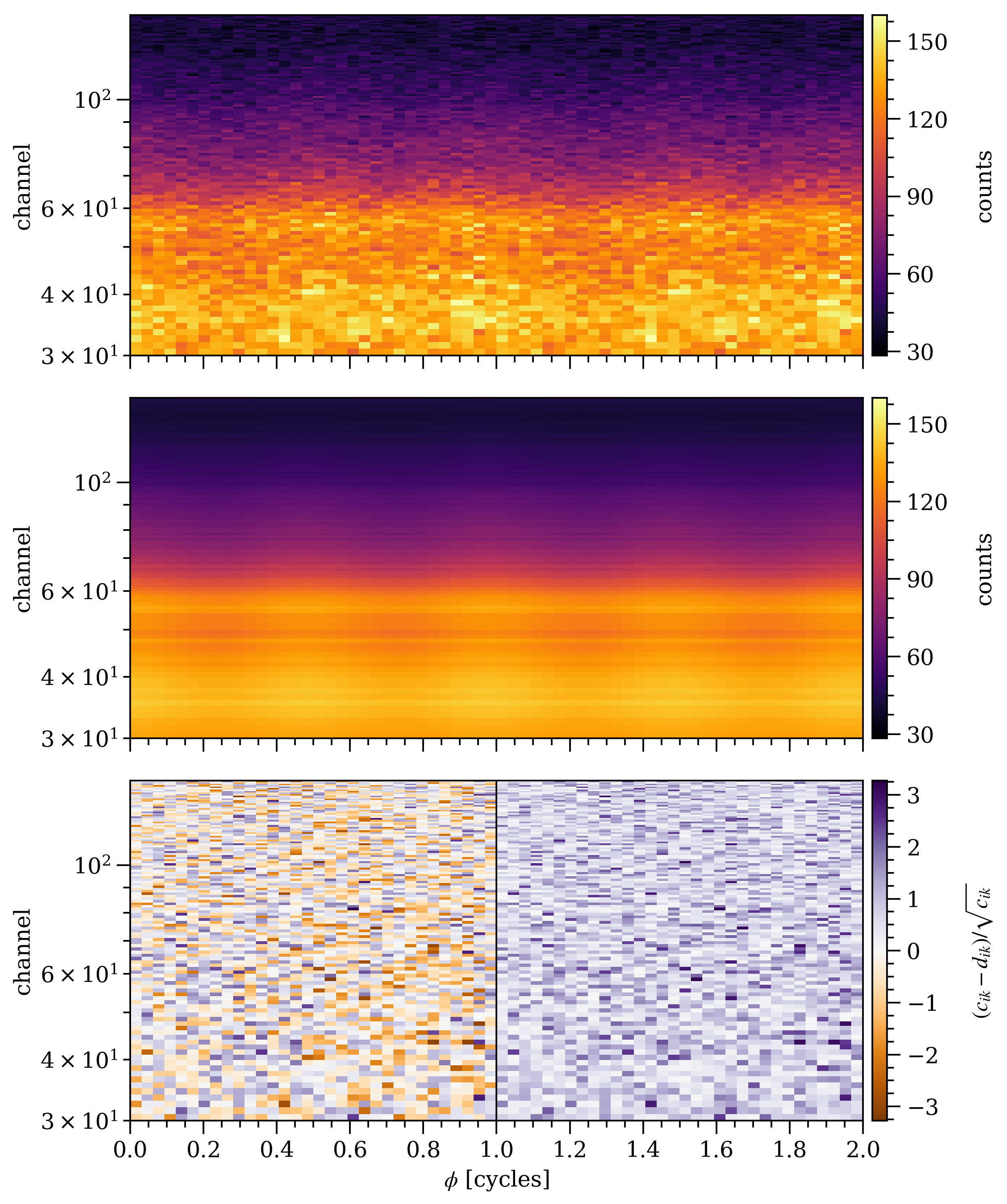}}    
    \caption{\small{
\NICER 3C50 count data, posterior-expected count numbers, and (Poisson) residuals for \texttt{ST-U} (left panel) and \texttt{ST-S} (right panel) conditional on 3C50 with ($n_{\mathrm{l}}=3$ and $n_{\mathrm{u}}=\infty$) background limits and \xmm likelihood function. 
See Figure 6 of \citetalias{Riley2021} for additional details about the figure elements. 
The complete figure set (3 images), which includes the corresponding \texttt{ST-Ua} residuals, is available in the online journal (HTML version). 
    }}
    \label{fig:resid0}
    \end{figure*}
}





{
    \begin{figure}[t!]
    \centering
    \resizebox{\hsize}{!}{\includegraphics[
    width=\textwidth]{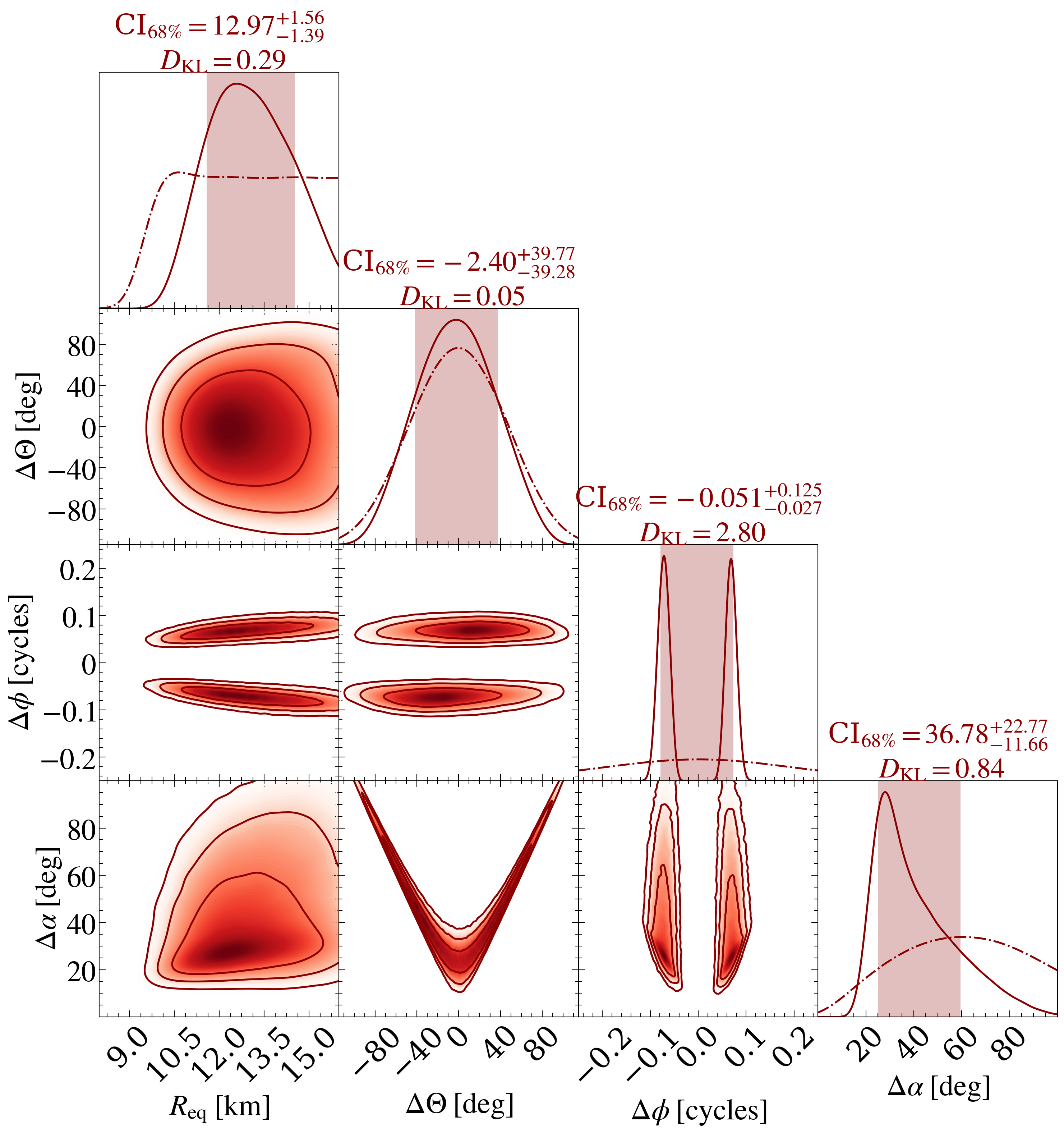}}
    \caption{\small{
    Antipodal offset inferred from the \texttt{ST-U} run based on 3C50 data with $n_{\mathrm{l}}=3$ background lower limit. 
    Here $\Delta \Theta$ is the offset from antipodality in the hot spot colatitudes, $\Delta \phi$ is the offset from antipodality in the spot azimuths, and $\Delta \alpha$ is the total offset angle from antipode. 
    See the caption of Figure \ref{fig:spacetime_SW} for additional details about the figure elements.  
    The complete figure set (2 figures), which includes the inferred offset also for the run with W21 data set and the smoothed SW background lower limit, is available in the online journal (HTML version).    
    }}
    \label{fig:post_antipoff}
    \end{figure}
}

{
    \begin{figure*}[t!]
    \centering
    \includegraphics[
    width=\textwidth]{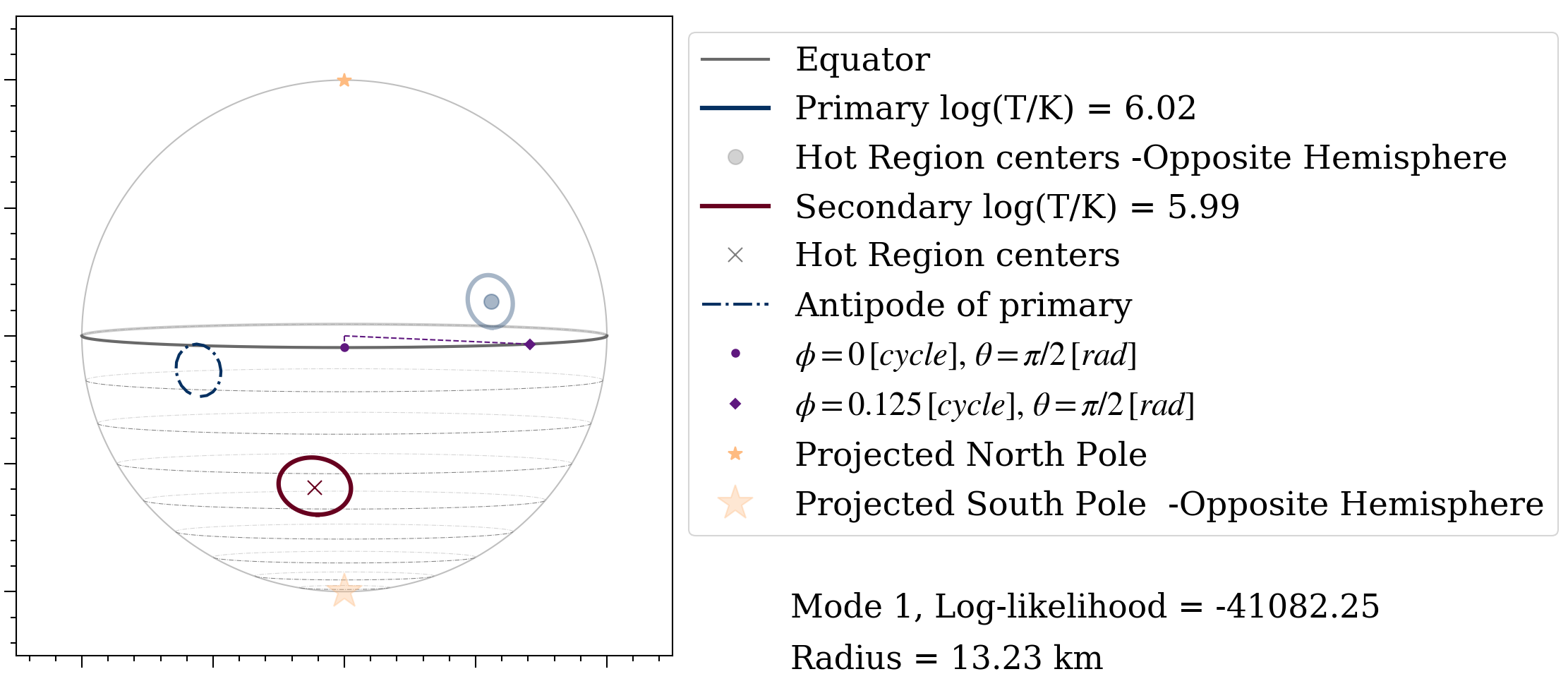}
    \caption{\small{
    Geometry configuration for STU-3C50-3X+XMM run (see Figures \ref{fig:spacetime_3C50XMM} and \ref{fig:spacetime_3C50_STS} for description) corresponding to the nested sample with maximum likelihood. 
    The viewing angle represents the Earth inclination to the spin axis.  
    The full animation set with 100 best samples, in the order of decreasing likelihood is shown in the online journal separately for the two different posterior modes (where the primary, i.e., lower colatitude spot is either around 0.4 cycle ahead or behind the secondary spot).  
    The animation shows a large variation between the best-fitting geometries. 
    More inferred configuration samples for different runs, also ordered based on maximum posterior weight instead of maximum likelihood, are available in \citet{salmi_zenodo22}. 
    }}
    \label{fig:geom_best}
    \end{figure*}
}

From the figures mentioned above, we see that the inferred radius is consistent with \texttt{ST-S} and \texttt{ST-U} (the former with notably larger credible intervals), but the \texttt{ST-Ua} model predicts a significantly higher radius.
However, both of the antipodal models show more residual clustering than \texttt{ST-U} (indicating defects in the model), even though the difference is not enormous.
The residuals for \texttt{ST-Ua} are only shown in the online figure set for Figure \ref{fig:resid0}, but they look very similar to those of \texttt{ST-S}. 
Also, the Bayesian evidence for \texttt{ST-U} is larger than the evidence for either of the antipodal models (see the evidences in Table \ref{table:3c50_STS}). 
On the other hand, the evidence for \texttt{ST-Ua} is higher than the evidence for \texttt{ST-S} when the background is not constrained  (and \texttt{ST-Ua} is clearly then predicting a background that is too low), and vice versa for the limited background case. 
This shows that, when the background is forced to be reasonable, adding the freedom of having different temperatures and spot sizes for the two hot regions is not helpful in order to resolve the structures in the residuals. 
Instead, \texttt{ST-Ua} tends to find significant differences in the temperature and size of the two spots, whereas \texttt{ST-U} (for \joh) finds very similar spot parameters but places the spots in a non-antipodal configuration. 

The major reason for antipodal configurations not being adequate for the modeling is the azimuthal offset of the two hot spots, rather than the offset in their colatitudes. 
The total angular offset of the secondary hot spot center from the antipode of the center of the primary spot one is inferred to deviate more than $25 ^\circ$ with more than $84 \%$ probability in the case of all  \texttt{ST-U} runs performed here (and in \citetalias{Riley2021}). 
The inferred angular sizes of the hot spots peak typically around $10 ^\circ$ with a tail in the posterior distribution toward larger spot sizes, being still less than $25 ^\circ$ with more than $84 \%$ probability (depending slightly on the model). 
This means there is only a relatively small chance that the antipode of the primary spot center would be enclosed by the secondary hot spot. 

In addition, we examined the effect of setting a hard upper limit of $10 ^\circ$ on the maximum angular size for the smaller of the hot spots.
This prior is somewhat more restrictive than expected from the maximum spot size from a centered-dipole \citep[see e.g.,][]{GonthierHarding1994}, but it was chosen to maximize the effect and to account for the fact that the entire polar cap should not actually be uniform in temperature. 
We tested this with low-resolution runs for 3C50 without any background constraints, because the inferred spot size is the highest for that case (the median of the smaller spot angular radius around $13 ^\circ.8$). 
We found that, by adding the spot size upper limit, the inferred radius shifted up roughly by 1 km, and the inferred background became also higher. 
Thus, having a spot size prior changed the results in the same direction as applying a lower bound to the background. 
However, this prior did not have any significant effects on the offset from antipodality of the spot centers, but of course the chance of the antipode being enclosed within the other spot became even smaller. 

The posterior distributions for the latitudinal, longitudinal, and total angular offset are shown in Figure \ref{fig:post_antipoff}. 
From there we see that colatitudes of the spots are inferred with large uncertainty, being on average closely antipodal. 
But the phase offset from the antipode is clearly constrained to around 0.1 of the total cycle, which is around $36 ^\circ$.  
The offset is also similar for both posterior modes, which only differ based on which of the spots (with higher or smaller colatitude) is leading the other spot by 0.4 cycle (these modes were also present in \citetalias{Riley2021}). 
The most likely spot configurations are visualized in the animation of 100 highest-likelihood geometries shown in the online journal version of Figure \ref{fig:geom_best}, with separate videos shown for the 2 different posterior modes.  

\section{Discussion}\label{sec:discussion}

As shown in Section \ref{sec:results}, we found that the results obtained when using \NICER background constraints in our analysis are consistent with those obtained previously from joint \NICER and \xmm analysis. 
In the following sections, we discuss the uncertainties in the measured NS radius in more detail, implications for the cross-calibration between \NICER and \xmm, and what we can learn from the inferred geometry of the radiating hot spots. 

\subsection{Radius Uncertainties}\label{sec:radius_uncertainties}

Using 3C50-selected \NICER data, we found larger statistical uncertainties in the inferred radius compared to the original run with the W21 data set. 
Applying the 3C50 background constraint improved the radius constraints slightly, the credible intervals becoming comparable to those obtained previously with the joint \NICER and \xmm analysis. 
On the other hand, applying the SW estimate to the W21 data set increased the statistical radius uncertainty, but again produced results closer to the joint \NICER and \xmm analysis.
This likely implies that the unmeasured systematic uncertainties are reduced when setting reasonable limits for the inferred background. 
We also found that the results are quite sensitive to the hard cutoffs set for the \NICER background prior information and to  small features in the background estimates (either by using a smoothed estimate or the mdb estimate that was forced to not exceed the data). 
All of the 68 \% credible intervals for the different runs presented here still overlap significantly. But using the new approach, i.e., constraining the \NICER background through the SW and 3C50 background estimates, the uncertainty associated with the cross-calibration between \NICER and \xmm is avoided. 

If we assume an unconstrained background, we find that the 3C50 data set produces a larger credible interval for the radius compared to W21. 
But after constraining the background using \xmm, both W21 and 3C50 produce comparably tight radius constraints. 
Statistical uncertainties depend on source and background counts \citep[see][]{lomiller13,Psaltis2014}. 
This implies a dependence of the radius credible interval on exposure time and effective area. 
Since 3C50 has both a shorter exposure time (1.55569 Ms versus 1.60268 Ms) and lower effective area (3C50 collects data from 50 of the 52 active detectors, W21 from 51 of them), the trend is consistent with expectations. 
However the difference between the statistical uncertainties of the two data sets (when not including \xmm) is bigger than expected by simply considering these two changes. 
In particular, based on the square root dependence on the observed counts, one would expect less than 0.1 km broadening in the length of the radius credible interval, instead of the observed broadening of 1 km. 
However, this is not necessarily surprising: the two data sets rely on two different observing periods and have indeed been built adopting very different approaches (3C50 targeting an estimation of the background and W21 targeting high counts) and procedures, including filtering. 
Because of the different observing periods, the contribution from a time variable background (like the active galactic nucleus, hereafter AGN, in the FoV) could also be different in the two data sets.

\subsection{Implications for the Cross-calibration Uncertainty between \NICER and \xmm}

Since the results of the new \NICER-only analysis are consistent with the joint \NICER and \xmm (using either old W21 or new 3C50 \NICER data sets), we conclude that there is no evidence for inconsistency in the \NICER and \xmm cross-calibration. 
This is also concluded in the independent analysis of Appendix \ref{apndx:A}. 
Instead, we found that the constraints obtained by applying \NICER background estimates are not restrictive enough to limit the inferred background as tightly as with a simultaneous \NICER and \xmm analysis. 
However, we note this is not necessarily the case for other sources and data sets, and the use of \NICER background estimates also provides an independent way to constrain the relative pulse amplitude for the signal from the hot spots, and thus more robustly measure the NS radius.
Also, as seen in \citetalias{Riley2021} and \citet{Miller2021}, the radius posteriors may slightly depend on the allowed energy-independent cross-calibration uncertainty in the modeling. 
In the new joint analysis of this paper, we primarily used the compressed scaling uncertainty from \citetalias{Riley2021}, and checked with a low-resolution run that further compression did not change the results significantly (as explained in Section \ref{sec:nicer_3C50_bkg_XMM}). 
However, energy-dependent cross-calibration effects may also exist, and those were not accounted for in our analysis.

\subsection{Deviation from Antipodality}

As presented in Section \ref{sec:nicer_STS}, we constrained the offset angle from antipodality to be more than $25 ^\circ$ with a higher than 84 \% probability even when allowing the possibility of the two spots having different temperatures and sizes. 
As shown in Table \ref{table:3c50_STS}, the evidence difference significantly favors the non-antipodal \texttt{ST-U} model both when including and excluding background prior information.
The Bayes factor of \texttt{ST-U} against the best antipodal model is always larger than $10^{8}$, which can be considered as decisive evidence against the antipodal models \citep{KassRaftery1995}. This result challenges magnetic field models with a centered-dipolar field structure, even though the inferred configuration is much closer to antipodal than in the case of \jdbl \citep{Bilous_2019,MLD_nicer19,Riley2019}. 
We also note that deviation arises almost entirely because of the phase difference of the hot spots, offset by 0.1 cycle from being antipodal. 
For this reason, the simpler models \texttt{ST-S} and \texttt{ST-Ua} explored in this paper were not able to provide as good a description of the data as the less constrained \texttt{ST-U} model. 
Since, in the \texttt{ST-U} runs, the inferred spot radii and temperatures are almost identical between the two spots, and colatitudes on average antipodal, it may be still possible to fit the data with a less complex model in which temperatures, sizes, and colatitudes are shared or derived between the spots but where each of them would have their own phase parameters. 

\section{Conclusions}
\label{sec:conclusions}

In this paper, we have presented PPM analysis of \NICER data of the high-mass pulsar \joh, using the \XPSI simulation and inference code, building on the work previously reported by \citetalias{Riley2021}, and using a similar framework as in \citet{Miller2021}, the latter reported in Appendix \ref{apndx:A}. 
Those previous analyses used \xmm data and background estimates to provide an indirect constraint on the \NICER background, and the inferred radius was sensitive to the inclusion of this constraint.  In this paper, we have used newly developed \NICER background models as a direct constraint during the PPM, so that we no longer have to attempt to model the uncertain cross-calibration between \NICER and \xmm.  The results, in particular the inferred mass and radius, are consistent with our previous findings and have similar uncertainties. This means that dense matter EOS inference using our previous results does not, at this stage, need to be updated; though the posterior samples derived from our new \NICER-only analysis can be downloaded via the associated repository in \citet{salmi_zenodo22}. 

Constraints are expected to become tighter in the future, as \NICER builds up more data on \joh.  The procedure outlined in this paper for analyzing \NICER data with \NICER-only background estimates will be used in future PPM analysis for both \joh and other \NICER sources. Our analysis shows the impact, on an inferred radius and its uncertainties, of having good and well-constrained models for the background emission. The 3C50 model (with background estimates and uncertainties) developed by \citet{remi22} used in this paper is extremely useful for PPM; in the future, we also hope to have better constraints on contributions from unrelated sources in the FoV such as the AGN SDSS J074115.14+662234.9, which as reported by \citet{Wolff20} also contributes to the \NICER background but is not accounted for in the 3C50 model estimates. Given the uncertainty over the AGN contribution at present, our recommendation for anyone wishing to use the updated \NICER-only posteriors reported in this paper is to use the 3C50-3X results, for which the inferred mass and radius are $2.073_{-0.066}^{+0.068}$ \msol and $12.97_{-1.39}^{+1.56}$ km.   

The consistency we found between the \NICER-only radius posteriors using \NICER backgrounds, and the radius posteriors from joint \NICER and \xmm analyses, motivates the inclusion of \xmm data in future work. 
Applying \NICER background information separately or on top of \xmm can still be a useful independent precaution. 
In our case, using both \NICER background estimates and \xmm reduced the uncertainty in the radius giving, for the 3C50-3X+XMM case; $M=2.075_{-0.067}^{+0.067}$ \msol and $R_{\mathrm{eq}}=12.90_{-0.97}^{+1.25}$ km. 
Furthermore, we note that the lower limit of 68 \% credible interval for the radius converges to very similar values also in the case of the corresponding Illinois-Maryland analyses (see Appendix \ref{apndx:A}, and Figure \ref{fig:radius_intervals} in Appendix \ref{apndx:tables}), when both 3C50 and \xmm data are applied, whereas a larger discrepancy appears for the joint W21 and \xmm analysis. 
This similarity can, at least partly, come from the improved data selection leading to smaller sensitivity on the background modeling. 
The difference in the 68 \% upper limit can still be attributed to the different choices in the radius prior and modeling procedures as discussed in Section 4.6 of \citet{Miller2021} and Section 4.4 of \citetalias{Riley2021}.

We also reported an extended analysis of the degree to which the inferred hot spot (magnetic polar cap) geometry deviates from a purely antipodal configuration. The degree to which the hot regions deviate from a purely antipodal configuration is of importance to those modeling multiwavelength pulsar emission and pulsar magnetic field evolution \citep{Bilous_2019,Chen2020,Kalapotharakos2021}. A purely antipodal configuration, even one where the hot regions can have different size and temperature, can be ruled out with a high degree of certainty.  The degree of offset of the central points of the polar caps is at least $25 ^\circ$ with 84 \% probability. 
This excludes, with high confidence, a centered-dipolar field configuration for \joh, implying that multipolar or non-centered-dipolar fields might be common in millisecond pulsars, as \joh is already the second object showing evidence for this. 
\\
\\
\begin{acknowledgments}
This work was supported in part by NASA through the \NICER mission and the Astrophysics Explorers Program. T.S., S.V., D.C., T.E.R., and A.L.W. acknowledge support from European Research Council (ERC) Consolidator grant (CoG) No.~865768 AEONS (PI: Watts).  
This work was sponsored by the Nederlandse Organisatie voor Wetenschappelijk Onderzoek (NWO) Domain Science for the use of the national computer facilities. 
A major part of the work was carried out on the HELIOS cluster including dedicated nodes funded via the abovementioned ERC CoG.  
S.G. acknowledges the support of the Centre National d'Etudes Spatiales (CNES). 
W.C.G.H. acknowledges support through grant 80NSSC22K0397 from NASA. 
S.M.M. thanks NSERC for research support. 
The material is based upon work supported by NASA under award number 80GSFC21M0002 (Z.W.). 
A.J.D. was supported in part by LANL/LDRD under project number 20220087DR. 
The Los Alamos Unlimited Release (LA-UR) number is LA-UR-22-28454. 
M.C.M. and A.J.D. were supported in part by NASA ADAP grant 80NSSC21K0649.  Part of this work was performed at the Aspen Center for Physics, which is supported by U.S. National Science Foundation grant PHY-1607611.  Some of the resources used in this work were provided by the NASA High-End Computing (HEC) Program through the NASA Center for Climate Simulation (NCCS) at Goddard Space Flight Center. 
NICER research at Naval Research Laboratory (NRL) is supported by NASA. 
A.L.W. would like to thank Anatoly Spitkovsky for prompting us to look in more detail at the question of non-antipodality. 

\end{acknowledgments}

\bibliographystyle{aasjournal}
\bibliography{allbib}

\clearpage 

\appendix

\section{3C50 Data and Background Model}\label{sec:3C50_data_and_bkg}

Here we describe both the data analysis steps and the filtering steps that define a set of GTIs that we refer to, hereafter, as the 3C50 data set. There is substantial overlap with other \NICER data sets for \joh, but the selection criteria are designed to produce a background spectrum with reliable uncertainty estimates, in any range of photon energy. 
We also note that the GTI sorting method from \citet{Guillot2019}, used to maximize the pulsed signal significance, was not used when filtering the new 3C50 data. Therefore, the new filtering can not produce any biases in the pulsation amplitude as suggested by \citet{Essick2022} for the original sorting method (see Section \ref{sec:model_sw}). 

The 3C50 background model \citep{remi22} uses three non-source count rates within the on-source GTIs to select and rescale background components from the model's libraries. The spectra in the libraries are derived from observations of seven fields (named ``BKGD\_RXTE") selected in the Rossi X-ray Timing Explorer era \citep{JMR06} to contain no detected sources with the existing instrumentation at that time.\footnote{We note that none of these seven blank-sky regions have been observed with modern X-ray imaging instruments (except eROSITA). There are no detected second ROSAT Source Catalog \citep[2RXS;][]{Boller2016}  sources within 8\arcmin\ of the NICER pointed positions of these fields, except for ''BKGD\_RXTE\_3'' which has two 2RXS sources (19 and 12 $\mathrm{c}~\mathrm{s}^{-1}$) within $\sim 4\arcmin$ and $\sim 4\arcmin.4$.}
These blank-sky observations are the same as used by the SW model in Section \ref{sec:model_sw}. 
They show that the in-band background rate, i.e., the rate of events that cannot be distinguished from X-ray events from the science target, can vary by several orders of magnitude, and they mainly originate from interactions between the silicon drift detectors and the various types of charged particles encountered in the space environment.

Appendix \ref{app:A1} describes the methods used to limit the systematic error during the initial data analyses and also to screen the background-subtracted spectra to avoid intervals when the background prediction has poor quality. Then in Appendix \ref{app:A2}, the large archive of final GTI selections for \joh is divided into eight consecutive time intervals, with exposure time $\sim 214$ ks in each interval. These data represent eight trials to measure the source intensity, choosing evaluations in three energy bands. The net count rate for \joh is 0.028 $\mathrm{c}~\mathrm{s}^{-1}$ at 0.3-2.0 keV, with no detection at 2-4 or 4-12 keV. Similar analyses are conducted for six additional rotation-powered ms pulsars, with the following conclusions. 
The mean and rms values are consistent with steady emission for each target, over the 5 yr lifetime of NICER, as expected for this source class. Furthermore, the rms measures in each energy band reveal common uncertainty values that scale proportionally with the shape of the average background spectrum. These rms results imply a systematic uncertainty that scales ($ 1 \sigma$) as roughly 2 \% of the average background spectrum, under the condition that the screening steps are imposed. 


\subsection{Data Processing and Filtering of Good Time Intervals}\label{app:A1}

For any \NICER target observed over different epochs of the data-processing and calibration pipeline, archived data must be brought to a uniform gain calibration by running the HEASARC tool $nicerl2$ on each observation. This task reruns the \NICER data pipeline to determine the time and equivalent photon energy of each recorded event, producing the 
$unfiltered$ and 
$cleaned$ event lists within GTIs that establish selection windows for the calibrated event lists. The pipeline uses an embedded task, $nimaketime$, to define GTIs while applying filters for a variety of geometric constraints, e.g., the pointing offset from the target, presence in the South Atlantic Anomaly, and minimum angles to the Earth limb and to the solar-illuminated Earth limb.  There are also rate constraints that specify maximum frequencies of overshoot-flagged events and undershoot-flagged events, and a maximum for the relationship between the overshoot rate and the magnetic shielding index, known as the cutoff rigidity, i.e., the COR\_SAX column in the \NICER filter files (see the documentation for ``nimaketime'' in the NASA HEASARC ftools web pages). In the present investigation, we adopted all of the default settings for geometric filters.  However, the three rate filters were effectively disabled by specifying impossibly high values for each (i.e., 15000) on the $nicerl2$ command line. The reason for this is to modestly increase the initial, overall exposure time, with the intent to filter for data quality at later stages of analysis.  

After reprocessing is finished, an explicit list of GTI times can be made by running \textit{nimaketime} on each observation, using the same combination of filter parameters, i.e., adopting default values for geometric parameters and disabling values for rate parameters. However, when defining our GTIs, two additional steps were employed to facilitate the background estimation process. 

The goal is to limit the dynamic range of values for the 3C50 model parameters within any GTI, so that non-linearities inherent in the background behavior might be reduced.  First, it was found that transitions in orbit day versus night can reduce the effectiveness of the $nz$ parameter (count rate at below 0.25 keV) for predicting the soft X-ray excess tied to optical light loading, when such transitions occur within a GTI.  Furthermore, the reflections in advance of ISS sunrise and lingering detector noise after ISS sunset can broaden the effects of the day-night transition to a range of $\pm 30$ s from the transition time.  To avoid integrating the $nz$ parameter over such conditions, $nimaketime$ was run twice with added constraints, SUNSHINE=1 and SUNSHINE=0, respectively, and then combining all of the results into one table.

A second step to improve background predictability is a response to the fact that \NICER background count rate and spectral shape routinely vary in complex ways over the course of the ISS orbit, and also from orbit to orbit. It is therefore prudent to limit the orbital phase duration of each time interval for which a background prediction will be made.  On the other hand, the 3C50 model parameters for the stage (1) library (i.e., $ibg$ and $hrej$; see Section 3.1 of \citealt{remi22} for definitions) suffer from low count rates, creating an opposing motivation to integrate as long as possible, to limit the effects of Poisson noise. Given the four passages between the Earth equator and the highest polar latitudes ($52 ^\circ$) in the ISS orbit (92 minutes), we choose a target and maximum GTI of 300 and 450 s, respectively.  Intervals of duration ($dt$) longer than 450 s are subdivided into $N$ GTIs, with $N = int (dt / 300 + 0.5)$, where $int$ denotes an integer value. Finally, we choose to ignore GTI when the gap time is 2 s or less.  Such gaps may be caused by packet losses or noise in $nimaketime$ selection parameters, and the isolated, brief gaps can be integrated over without significant consequences.  To implement our choices for GTI definition, we use a customized C program to read the initial GTI table and then output the final table, while ignoring brief gaps, masking mission elapsed time values within 30 s of a SUNSHINE transition in consecutive GTIs, and dividing long GTIs into pieces of roughly 300 s.  The final GTI table is indexed, and the index number is included in the file names of all downstream data products based on a given GTI. 

For \joh, we applied these steps to \NICER data collected through between 2018 September 21 and 2021 December 28 (observation IDs, hereafter ObsIDs, 1031020101 through 4031020407), yielding 9907 GTIs and a total exposure of 2.79 Ms.  This is the starting point for application of the 3C50 background model.  For every GTI, the extractions of spectra and 3C50 model parameters values are made on the basis of 50 selectable FPMs, while ignoring FPMs 14 and 34, as explained in \citet[][in contrast, the W21 data set events did not exclude detector 14]{remi22}. 
The model parameter values are used to generate a 3C50 prediction of the background spectrum, again per GTI, and the difference between the raw extraction and the 3C50 background spectrum is the background-subtracted spectrum used for subsequent filtering considerations.

Screening criteria use background-subtracted spectra (noted as parameters with subscript ``net''), per GTI, in diagnostic energy bands, as described in \cite{remi22}. The criteria are designed to flag and ignore the GTIs that have an obviously poor background prediction. For faint, non-accreting pulsars, such as \joh, we extend the types of filters in the interest of reducing systematic uncertainty in the background estimation. The list of filters and the fraction of remaining exposure time, in parentheses, for the case of \joh, are given as follows.  The selection fractions are scaled to the initial set of GTIs with a total exposure 2.79 Ms.  We first remove GTIs that have 3C50 parameter measurements beyond the model limits (0.998). Then, for better 3C50 parameter statistics, we select GTIs with exposure time greater than 200 s (0.968). To avoid noise contamination in strong sunlight, we select GTIs with $nz < 220$ $\mathrm{c}~\mathrm{s}^{-1}$ (0.856). Because the 3C50 model performs best at normal-low count rates, presumably due to the lack of identified metrics for background contributions from trapped and precipitating particles in the local space environment, we further select GTIs with $ibg < 0.2$ $\mathrm{c}~\mathrm{s}^{-1}$ (0.790).  Extending the filters described in \cite{remi22}, we apply $|hbg_{\mathrm{net}}| < 0.05$ (0.777) and $|S0_{\mathrm{net}}| < 0.15$ (0.728). 

Consistent with the overall result that there is no detection of \joh with \NICER above 2.0 keV, we use the in-band energy ranges at 2-4 keV ($C$ band) and 4-12 keV ($D$ band) to select $|D_{\mathrm{net}}| < 0.3$ (0.686) and  $|C_{\mathrm{net}}| < 0.1$ (0.614). 
The use of  $D_{\mathrm{net}}$  and $C_{\mathrm{net}}$  filters has to be self-consistent with the final, deep FoV spectrum of a given target. Filter limits were chosen to reject the wings of the probability distribution in each of these bands, assuming that such outliers arise from inaccurate background predictions. 
If the filter limits are within an order of magnitude of the net flux in that band, then the filtering step is suspended. 
Finally, for the simplicity of using one response file for these data, we select from the remainder only those GTIs in which all 50 selected FPMs are operating (0.558). This completes the definition of the 3C50 data set (1.55569 Ms final exposure) used in this investigation. The source intensity is 0.028 $\mathrm{c}~\mathrm{s}^{-1}$ at 0.3-2.0 keV, and there is no detection at 2-4 or 4-12 keV.

The sum of the raw spectra for the 3C50 data set and the sum of the 3C50 model background predictions for the same time intervals are shown in Figure \ref{fig:W21_vs_3C50_data}. The difference between raw and background spectra is referred to as the average FoV spectrum for \joh, and the possibility of contributions from other sources in the \NICER FoV must be considered before treating the FoV spectrum as the average pulsar spectrum. The same time intervals for GTIs in the 3C50 data set were used to derive the average 3C50 pulse profile for \joh, using the same steps applied to other data sets.

\subsection{Quantifying Systematic Uncertainty in the Background Spectrum}\label{sec:3c50_uncertainties}\label{app:A2}
 
The group of non-accreting X-ray pulsars with large exposure times in the \NICER archive (currently, 1-3 Ms per target) offers an important opportunity to measure the systematic limits on the accuracy of \NICER background models. The ability to assume that the average source intensity is constant over the timescale of many years is another key condition.  For each target, we divide the exposure time into subsets of 100-200 ks, average the background-subtracted spectra for each subset, and finally use the rms scatter for each subset group, in selected energy bands, to search for evidence of systematic limits that span the group of such pulsars.  We stress that the average background spectrum is always tied to a specific list of GTIs, with substantial variations that depend on data selection criteria.  Background measurements and model results, alone, do not offer a control needed to anchor uncertainty analyses.  On the other hand, deep background-subtracted spectra for invariable X-ray sources, examined as a series of 100+ ks exposures with negligible statistical error, might reveal systematic limits in the background model, given a uniform set of filtering steps. The net count rate for such pulsars is fainter than the average background level, giving rise to the expectation that systematic uncertainties would produce the same rms variations in 100-200 ks subsets, for a given energy band, regardless of the pulsar brightness. However, we note that the assumption for this method is that there are no variable sources in the FoV. 

Filtering efforts were conducted for 7 pulsars with Ms accumulated exposures in the \NICER archive, and the results are given in Table~\ref{tab:3c50}.  For each pulsar, the data were divided into a number of subsets (column (6) in Table~\ref{tab:3c50}), yielding 100-200 ks exposures in each sample. The selection steps are the same as that described above for \joh, except for two differences.  For the tabulated results, the last step, i.e., the restriction to a uniform set of 50 active FPMs, was not imposed, while the spectra are linearly scaled to 50 FPMs for the small percentage of GTIs where one or more normally selected FPMs were not operating.  The selected exposure time for \joh is then 1.71 Ms, rather than the 1.556 Ms noted for data used in the present investigation. Despite this difference, the average net count rates for \joh in the 3 given energy bands (columns (7)-(9) in Table~\ref{tab:3c50}) are the same, in either case, for the number of significant digits reported.  The second difference in selection steps is that B1821-24 was not filtered for the $C$ band (2.4 keV, column (8) in Table~\ref{tab:3c50}), where a significant detection is found. This result can be seen as a corroboration of the filtering methods, since B1821-24 is a relatively bright pulsar with a non-thermal spectrum that continues above 2 keV \citep{rowa20}. The net count rates (columns 7-9 in Table~\ref{tab:3c50}) are background-subtracted averages and rms values for the given range in photon energy, in units counts per second, scaled to 50 FPMs. For example, we note that the C-band count rate for \joh is 0.028 c s$^{-1}$.

The rms values shown in Table~\ref{tab:3c50} are similar across all pulsars, for each of the chosen energy bands. This invites interpretation as a measurement of systematic limits for the accuracy of 3C50 model predictions, for 100-200 ks exposure times, given the filtering efforts described herein. We adopt the average rms values in counts per second, as estimates for the 3C50 background uncertainty (1 $\sigma$): 0.018 (0.3-2.0 keV), 0.007 (2-4 keV) and 0.010 (4-12 keV). The ratios in these energy bands scale as:  1, 0.39, 0.56. On the other hand, the background spectrum for the 1.55569 Ms data set for \joh shows integrated values of 0.4207, 0.1246, and 0.2401 c s$^{-1}$, in the same energy bands, scaling as follows: 1.0, 0.30, 0.57. This implies that the spectral shapes of the rms measurements for pulsar subsets and the background spectrum for \joh are quite similar. We interpret this as evidence that the systematic uncertainty in the 3C50 model can be pictured as a normalization uncertainty on the final background spectrum. The uncertainty array (versus spectral bin), for the purposes of pulsar pulse-phase modeling, is then estimated as the rescaled values (by factor 0.018 / 0.4207) of the 3C50 background spectrum for \joh.

\begin{deluxetable}{cccllcccc}
\tablecaption{Pulsar Data Sets Filtered for 3C50 Background Model}
\tablewidth{0pt}
\decimalcolnumbers
\startdata
\tablehead{ \colhead{Pulsar} & \colhead{ObsID} & \colhead{ObsID} & \colhead{Initial} & \colhead{Selected} & \colhead{Subset} & \colhead{0.3-2.0 keV} & \colhead{2-4 keV} & \colhead{4-12 keV} \\
            \colhead{Name} & \colhead{Start} & \colhead{End}     & \colhead{Ms}     & \colhead{Ms}      & \colhead{ }   & \colhead{Rate (rms) $^{\mathrm{a}}$} & \colhead{Rate (rms)} & \colhead{Rate (rms)} }
J0030+0451 &  1060020101  &  4060020511  &     3.078    &     1.984    &    9  &  0.234 (0.018) & -0.0077 (0.005) & -0.0015 (0.010) \\
J0614-3329 &  1030050106  &  5030050307  &     1.250    &      0.726    &    6  &  0.062 (0.016) & -0.0112 (0.008) & -0.0100 (0.011) \\
J0740+6620 &  1031020101  &  4031020407  &     2.790    &     1.713    &    8  &  0.028 (0.020) & -0.0042 (0.010) & -0.0061 (0.011) \\
J1231-1411 &  1060060101  &  5060060656  &     3.130    &     1.730    &    8  &  0.163 (0.019) & -0.0131 (0.006) & -0.0094 (0.008) \\
J1614-2230 &  0060310101  &  4060310289  &     1.049    &      0.572    &    5  &  0.216 (0.018) & -0.0030 (0.003) &  0.0063 (0.007) \\
B1821-24 $^{\mathrm{b}}$  &  1070010101  &  5070010449  &     1.139    &      0.789    &    6  &  0.666 (0.020) &  0.0747 (0.008) &  0.0236 (0.010) \\
B1937+21   &  1070020101  &  5070020735  &     2.562    &     1.079    &    6  & -0.085 (0.015) &  0.0044 (0.007) & -0.0013 (0.012) \\
\enddata
\tablecomments{\ \\ $^{\mathrm{a}}$ Rate is measured in counts per second. \\ $^{\mathrm{b}}$ B1821-24 was not filtered for the 2-4 keV band.}
\end{deluxetable}\label{tab:3c50}

\clearpage

\section{An Independent Analysis of the 3C50 Data Set and Background Estimates}\label{apndx:A}
Herein we describe analyses of the 3C50 data set and associated background estimates using the same PPM code and analysis pipeline as in \citet{Miller2021}, which we refer to subsequently as the Illinois-Maryland analyses. We investigate all combinations of inclusion and exclusion of both XMM data and NICER background estimates in conjunction with the 3C50-selected NICER data set. We provide an independent confirmation that the radius inferences from 3C50 data and background estimates are consistent with previous analyses of NICER and XMM data, and that the calibrations of NICER and XMM do not appear inconsistent with one another.

\subsection{Methodology}
We utilize the PPM models and code previously applied to PSR J0740+6620 in \citet{Miller2021}, the details and underlying assumptions of which are further described in \citet{MLD_nicer19}, \citet{Bogdanov2019b}, and \citet{Bogdanov2021}. As in \citet{Miller2021}, we perform Bayesian parameter inference using the parallel-tempered Markov Chain Monte Carlo (MCMC) sampler in version 2.2.1 of the emcee package \citep{emcee}.
We drew the initial walker positions for each analysis from a kernel density estimate of the posterior probability distributions from the analyses of \citet{Miller2021}, using the samples archived in \citet{miller_m_c_2021_4670689}.
We use the termination criterion described in Section 4.3 of \citet{Miller2021}, accruing $10^7$ post-convergence samples during each analysis. 

When including 3C50 background estimates, we first produce a smoothed model count spectrum using a cubic smoothing spline, smoothed such that the spline has a frequency response of 0.5 over a wavelength of 10 bins following the algorithm introduced by \citet{10150/261038}. To take into account systematic field-to-field uncertainty in the 3C50 background predictions, we introduce an energy-independent rescaling factor $\beta_{\rm 3C50}$ in the Illinois-Maryland analyses, by which we multiply the 3C50 background spectrum. Based on Appendix \ref{sec:3c50_uncertainties},  we use for this parameter a Gaussian prior with a mean of 1 and a standard deviation of 0.0461, a conservative estimate of the total uncertainty. 
On top of the smoothed and rescaled background spectrum predicted by the 3C50 model, we marginalize over additional possible background contributions (such as other X-ray sources within NICER's FoV; see \citealt{Wolff20}) as described in Section 3.4.1 of \citet{Miller2021}.
In all analyses using XMM data, we used the nominal telescope response, without allowing for any variation in the effective area \citep[see, for comparison, Section 4.4 of][]{Miller2021}. 

\subsection{Results}
\begin{figure}
\centering
\includegraphics[]{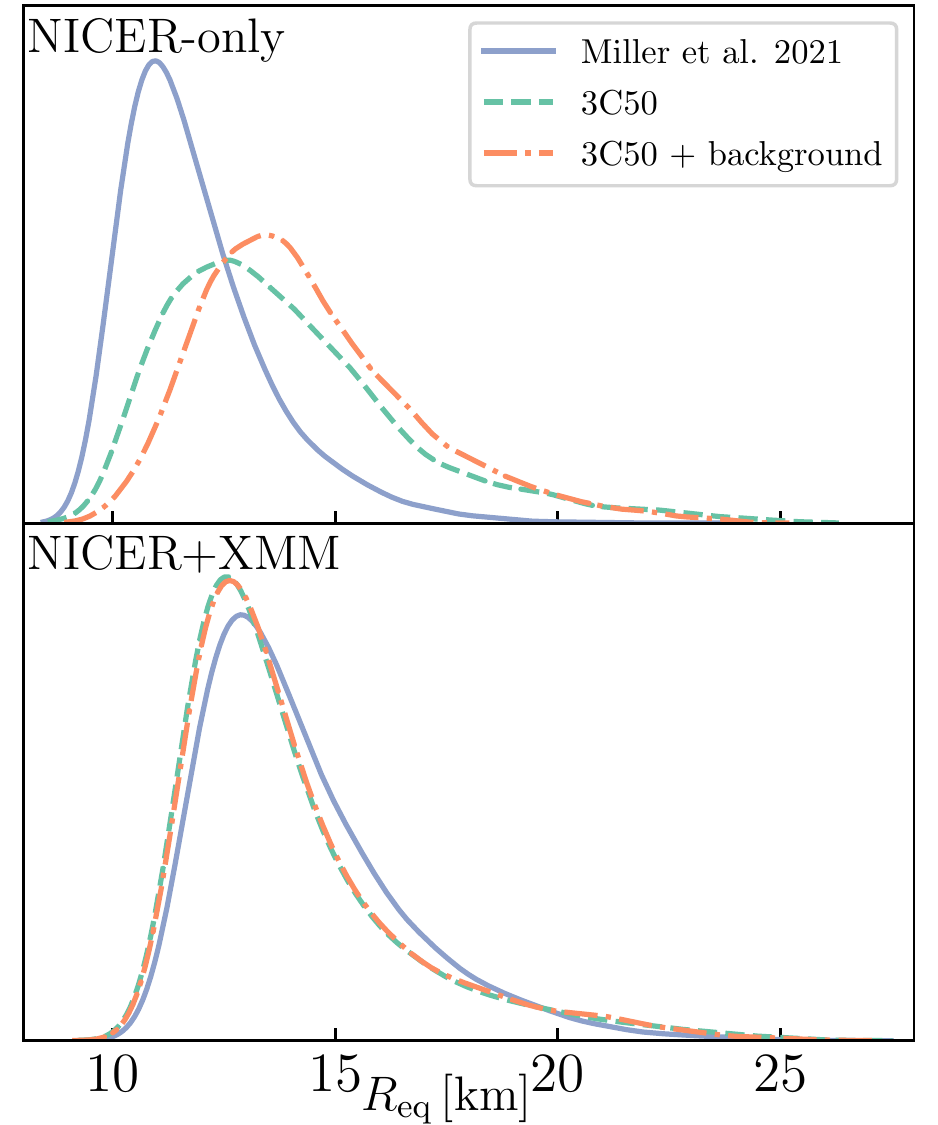}
\caption{The 1D posterior probability density of the equatorial circumferential radius of PSR J0740-6620 inferred using the Illinois-Maryland analysis pipeline. Each line is an averaged shifted histogram \citep[e.g.,][]{10.2307/2241123} of the posterior samples from the converged portion of an MCMC analysis. Blue solid lines illustrate the results from \citet{Miller2021}, which analyzed the W21 data set. Green dashed lines illustrate the results from analyses that used the 3C50-selected NICER data, but did not incorporate the 3C50 background estimate as a lower limit; and orange dashed-dotted lines illustrate results from analyses, which used the 3C50-selected NICER data set and incorporated the 3C50 background estimate, with uncertainties, as a lower limit on the NICER background. The top panel includes results from analyses of NICER data alone, while the bottom panel includes results derived using both NICER and XMM data.}
\label{fig:rcompIM}
\end{figure}

We present the inferred radii from the Illinois-Maryland analyses of the available X-ray data for PSR J0740+6620 in Figure \ref{fig:rcompIM}. The maximum-likelihood, median, and $68\%$ credible interval equatorial radius values derived from each analysis are presented in Tables \ref{tab:NICERonly} and \ref{tab:NICERXMM} for our NICER-only and NICER+XMM analyses respectively. Some differences between the X-PSI and Illinois-Maryland results are expected based on a number of choices made over the course of each set of analyses, including the details of the 3C50 background implementation in addition to those described in Section 4.6 of \citet{Miller2021} and Section 4.4 of \citetalias{Riley2021}. However, there are a number of common trends between the results reported in this Appendix and those derived using the X-PSI framework. 

In the Illinois-Maryland analyses not incorporating XMM data, the results of which are shown in the upper panel of Figure \ref{fig:rcompIM}, we find that inclusion of the 3C50 data alone pushes the inferred equatorial radii to higher values, relative to the NICER-only analysis of the W21 data, into closer agreement with the results from the NICER+XMM fit to the W21 data. This shift is likely the result of the more stringent data selection cuts when constructing the 3C50 data set, described in Appendix \ref{sec:3C50_data_and_bkg}, resulting in a lower background and thus a higher fraction of modulated emission, and is similar to the results presented in Figure \ref{fig:spacetime_3C50}. Furthermore, inclusion of background constraints in the fits only to NICER data shifts the equatorial radius posterior to yet higher values, a similar effect to including XMM data in the analysis of the W21 data set, along the same lines as the results discussed in Section \ref{sec:nicer_3C50_bkg_3C50}. As in Section \ref{sec:nicer_3C50}, we find that the posteriors derived from the 3C50 NICER data set, without incorporating XMM data, are broader than those derived from the W21 data set.

In the Illinois-Maryland analyses, which incorporate both NICER and XMM data, the results of which are presented in the lower panel of Figure \ref{fig:rcompIM}, we observe that including 3C50 background constraints makes a much smaller difference than in the NICER-only case, similar to the results presented in Section \ref{sec:nicer_3C50_bkg_XMM}. However, we do find that including both XMM data and a 3C50-derived NICER background estimate results in a very slight shift of the equatorial radius posterior to higher values. As shown in Table \ref{tab:NICERXMM} the $68\%$ credible regions become very slightly wider in both analyses of the 3C50 data set compared to the results from \citet{Miller2021}, although to a lesser extent than in the analogous NICER-only analyses.

\begin{deluxetable}{ccccc}
\caption{Best-fit, Median, and $\pm 68.3\%$ Credible Interval Values ($\pm\textrm{CI}_{68\%}$) of the Equatorial Circumferential Radius (Given in Kilometers) of PSR J0740+6620 from the Analysis of Various Data Sets, Using Only NICER Data}\label{tab:NICERonly}
\tablehead{
\colhead{NICER Data Set} & \colhead{Best Fit} & \colhead{Median} &
\colhead{$-\textrm{CI}_{68\%}$} &
\colhead{$+\textrm{CI}_{68\%}$}
}
\startdata
W21 & 11.008 & 11.512 & 10.381 & 13.380 \\
3C50 & 10.265 & 13.383 & 11.350 & 16.354 \\
3C50+bkg & 12.849 & 13.959 & 12.097 & 16.747
\enddata
\end{deluxetable}

\begin{deluxetable}{ccccc}
\caption{Best-fit, Median, and $\pm 68.3\%$ Credible Interval Values ($\pm\textrm{CI}_{68\%}$) of the Equatorial Circumferential Radius (Given in Kilometers) of PSR J0740+6620 from the Analysis of Various Data Sets, Using Both NICER and XMM Data}
\label{tab:NICERXMM}
\tablehead{
\colhead{NICER Data Set} & \colhead{Best Fit} & \colhead{Median} &
\colhead{$-\textrm{CI}_{68\%}$} &
\colhead{$+\textrm{CI}_{68\%}$}
}
\startdata
W21 & 13.823 & 13.713 & 12.209 & 16.326 \\
3C50 & 11.933 & 13.310 & 11.940 & 16.107 \\
3C50+bkg & 12.918 & 13.362 & 11.980 & 16.177
\enddata
\end{deluxetable}

\clearpage

\section{Mass and Radius Credible Intervals for Different Runs Using X-PSI}\label{apndx:tables}

We summarize the inferred results of radius and mass credible intervals using the \XPSI framework using different approaches for the background modeling in Tables \ref{table:sw} - \ref{table:3c50_STS}. 
The radius intervals are also visualized in Figure \ref{fig:radius_intervals} for both \XPSI and Illinois-Maryland results. 

\begin{longtable*}{l|lll|lll|lll|lll}
\caption{Summary Table for Different Runs with W21 NICER Data Set Including SW Background Estimate without Smoothing and with Smoothing}\label{table:sw}\\
\hline\hline
Parameter &  \multicolumn{3}{c|}{W21} & \multicolumn{3}{c|}{W21+XMM} &
\multicolumn{3}{c|}{W21-0.9xSW} & \multicolumn{3}{c}{W21-0.9xSWs} \\
\hline
$M$ $[M_{\odot}]$ &
$2.078_{-0.063}^{+0.066}$ &
$0.01$ &
$2.125$  &
$2.072_{-0.067}^{+0.066}$ &
$0.01$ &
$2.070$   &
$2.072_{-0.068}^{+0.066}$ &
$0.01$ &
$1.99$ &
$2.072_{-0.066}^{+0.067}$ &
$0.01$ &
$1.99$ \\
\hline
$R_{\textrm{eq}}$ $[$km$]$ &
$11.29_{-0.81}^{+1.20}$ &
$0.72$ &
$10.90$ &
$12.39_{-0.98}^{+1.30}$ &
$0.58$ &
$11.02$ &
$12.53_{-1.10}^{+1.47}$ &
$0.45$ &
$11.78$ &
$11.90_{-0.95}^{+1.32}$ &
$0.58$ &
$11.26$ \\
\hline\hline
\end{longtable*}
\tablecomments{We show the $68.3\%$ credible intervals around the median $\widehat{\textrm{CI}}_{68\%}$, Kullback–Leibler divergence $D_{\textrm{KL}}$ representing prior-to-posterior information gain, and the maximum likelihood nested sample $\widehat{\textrm{ML}}$ of mass and radius. See Figure \ref{fig:spacetime_SW} for more details.}

\begin{longtable*}{l|lll|lll|lll|lll}
\caption{Similar to Table \ref{table:sw}, but with Different 3C50 Runs using Smoothed Background Spectra} \label{table:3c50}\\
\hline\hline
Parameter &  \multicolumn{3}{c|}{3C50} & \multicolumn{3}{c|}{3C50-3X} & \multicolumn{3}{c|}{3C50-33} & \multicolumn{3}{c}{3C50-23} \\
\hline
\endfirsthead
\multicolumn{6}{c}%
{\tablename\ \thetable\---\textit{Continued from previous page}} \\
\hline
\endhead
\hline \multicolumn{6}{r}{\textit{Continued on next page}} \\
\endfoot
\hline\hline
\endlastfoot
\hline
$M$ $[M_{\odot}]$ &
$2.074_{-0.065}^{+0.067}$ &
$0.01$ &
$2.061$  &
$2.073_{-0.066}^{+0.068}$ &
$0.01$ &
$2.101$ &
$2.076_{-0.067}^{+0.067}$ &
$0.01$ &
$2.153$ &
$2.076_{-0.067}^{+0.068}$ &
$0.0$ &
$2.071$ \\
\hline
$R_{\textrm{eq}}$ $[$km$]$ &
$12.50_{-1.58}^{+1.69}$ &
$0.15$ &
$13.35$ &
$12.97_{-1.39}^{+1.56}$ &
$0.29$ &
$12.66$ &
$12.09_{-1.08}^{+1.41}$ &
$0.45$ &
$12.64$ &
$13.05_{-1.11}^{+1.36}$ &
$0.50$ &
$13.66$ \\
\hline\hline
\end{longtable*}
\tablecomments{See Figure \ref{fig:spacetime_3C50bkg} for the model name descriptions.}

\begin{longtable*}{l|lll|lll|lll|lll}
\caption{Similar to Table \ref{table:3c50}, but with Using the Minimum Function for the Lower Bound of the Background}\label{table:3c50mdb}\\
\hline\hline
Parameter &  \multicolumn{3}{c|}{3C50} & \multicolumn{3}{c|}{3C50-3X-mdb} & \multicolumn{3}{c|}{3C50-33-mdb} & \multicolumn{3}{c}{3C50-23-mdb} \\
\hline
\endfirsthead
\multicolumn{6}{c}%
{\tablename\ \thetable\---\textit{Continued from previous page}} \\
\hline
\endhead
\hline \multicolumn{6}{r}{\textit{Continued on next page}} \\
\endfoot
\hline\hline
\endlastfoot
\hline
$M$ $[M_{\odot}]$ &
$2.074_{-0.065}^{+0.067}$ &
$0.01$ &
$2.061$  &
$2.074_{-0.066}^{+0.067}$ &
$0.01$ &
$1.974$ &
$2.077_{-0.066}^{+0.067}$ &
$0.01$ &
$2.112$ &
$2.076_{-0.067}^{+0.067}$ &
$0.0$ &
$1.909$ \\
\hline
$R_{\textrm{eq}}$ $[$km$]$ &
$12.50_{-1.58}^{+1.69}$ &
$0.15$ &
$13.35$ &
$12.69_{-1.46}^{+1.66}$ &
$0.23$ &
$12.76$ &
$11.73_{-1.05}^{+1.39}$ &
$0.48$ &
$10.64$ &
$12.35_{-1.03}^{+1.38}$ &
$0.51$ &
$13.11$ \\
\hline\hline
\end{longtable*}
\tablecomments{See second figure in Figure set \ref{fig:spacetime_3C50bkg} for more details.}

\begin{longtable*}{l|lll|lll|lll|lll}
\caption{Similar to Table \ref{table:3c50}, but with joint 3C50 and \xmm Runs}\label{table:3c50_XMM}\\
\hline\hline
Parameter &  \multicolumn{3}{c|}{3C50+XMM} & \multicolumn{3}{c|}{3C50-3X+XMM} & \multicolumn{3}{c|}{3C50-33+XMM} & \multicolumn{3}{c}{3C50-23-XMM} \\
\hline
\endfirsthead
\multicolumn{6}{c}%
{\tablename\ \thetable\---\textit{Continued from previous page}} \\
\hline
\endhead
\hline \multicolumn{6}{r}{\textit{Continued on next page}} \\
\endfoot
\hline\hline
\endlastfoot
\hline
$M$ $[M_{\odot}]$ &
$2.075_{-0.067}^{+0.067}$ &
$0.01$ &
$2.119$ &
$2.075_{-0.067}^{+0.067}$ &
$0.01$ &
$2.132$ &
$2.076_{-0.067}^{+0.067}$ &
$0.01$ &
$1.972$ &
$2.075_{-0.067}^{+0.067}$ &
$0.0$ &
$1.934$ 
\\
\hline
$R_{\textrm{eq}}$ $[$km$]$ &
$12.89_{-0.97}^{+1.26}$ &
$0.61$ &
$13.59$ &
$12.90_{-0.97}^{+1.25}$ &
$0.62$ &
$13.23$ &
$12.71_{-0.94}^{+1.25}$ &
$0.64$ &
$12.00$ &
$12.88_{-0.95}^{+1.25}$ &
$0.64$ &
$12.80$ 
\\
\hline\hline
\end{longtable*}
\tablecomments{See Figure \ref{fig:spacetime_3C50XMM} for more details.}

\clearpage

\begin{longtable*}{l|lll|lll|lll|lll}
\caption{Similar to Table \ref{table:3c50}, but with Different Runs Using Antipodal Hot Spots}\label{table:3c50_STS} \\
\hline\hline
Parameter &  \multicolumn{3}{c|}{STS-3C50} & \multicolumn{3}{c|}{STUa-3C50} & \multicolumn{3}{c|}{STS-3C50-3X+XMM} & \multicolumn{3}{c}{STUa-3C50-3X+XMM} \\
\hline
\endfirsthead
\multicolumn{6}{c}%
{\tablename\ \thetable\---\textit{Continued from previous page}} \\
\hline
\endhead
\hline \multicolumn{6}{r}{\textit{Continued on next page}} \\
\endfoot
\hline\hline
\endlastfoot
\hline
$M$ $[M_{\odot}]$ &
$2.076_{-0.068}^{+0.070}$ &
$0.0$ &
$2.166$ &
$2.054_{-0.066}^{+0.071}$ &
$0.08$ &
$2.138$ &
$2.082_{-0.069}^{+0.070}$ &
$0.0$ &
$2.314$ &
$2.085_{-0.070}^{+0.068}$ &
$0.01$ &
$1.926$ 
\\
\hline
$R_{\textrm{eq}}$ $[$km$]$ &
$11.73_{-1.45}^{+2.18}$ &
$0.14$ &
$13.45$ &
$15.57_{-0.59}^{+0.31}$ &
$2.13$ &
$15.95$ &
$13.11_{-1.37}^{+1.73}$ &
$0.33$ &
$15.12$ &
$14.65_{-1.84}^{+0.97}$ &
$0.59$ &
$15.93$ 
\\
\hline
$\widehat{\ln\mathcal{Z}}$&
\multicolumn{3}{c|}{$-16024.67\pm0.02$} &
\multicolumn{3}{c|}{$-16018.03\pm0.02$} &
\multicolumn{3}{c|}{$-20593.12\pm0.02$} &
\multicolumn{3}{c}{$-20597.28\pm0.02$} 
\\
\hline\hline
\end{longtable*}
\tablecomments{We also show the log-evidences $\widehat{\ln\mathcal{Z}}$ $^{\mathrm{a}}$ to compare the model performance between \texttt{ST-U}, \texttt{ST-Ua} and \texttt{ST-S}$^{\mathrm{b}}$. For more details see Figure \ref{fig:spacetime_3C50_STS} and Section \ref{sec:nicer_STS}. \\ 
$^{\mathrm{a}}$ The evidence is defined as the prior predictive probability, e.g. $p(d_{\rm N}, d_{\rm X},\{\mathcal{B}_{\rm N}\},\{\mathscr{B}_{\rm X}\}\,|\,\TT{ST-S})$. Note that the evidences for models with different background limits are not comparable. \\ 
$^{\mathrm{b}}$ For STU-3C50 $\widehat{\ln\mathcal{Z}} = -15997.60\pm0.02$, and for STU-3C50-3X+XMM, $\widehat{\ln\mathcal{Z}} = -20564.85\pm0.02$.}




{
    \begin{figure*}[t!]
    \centering
    \includegraphics[
    width=\textwidth]{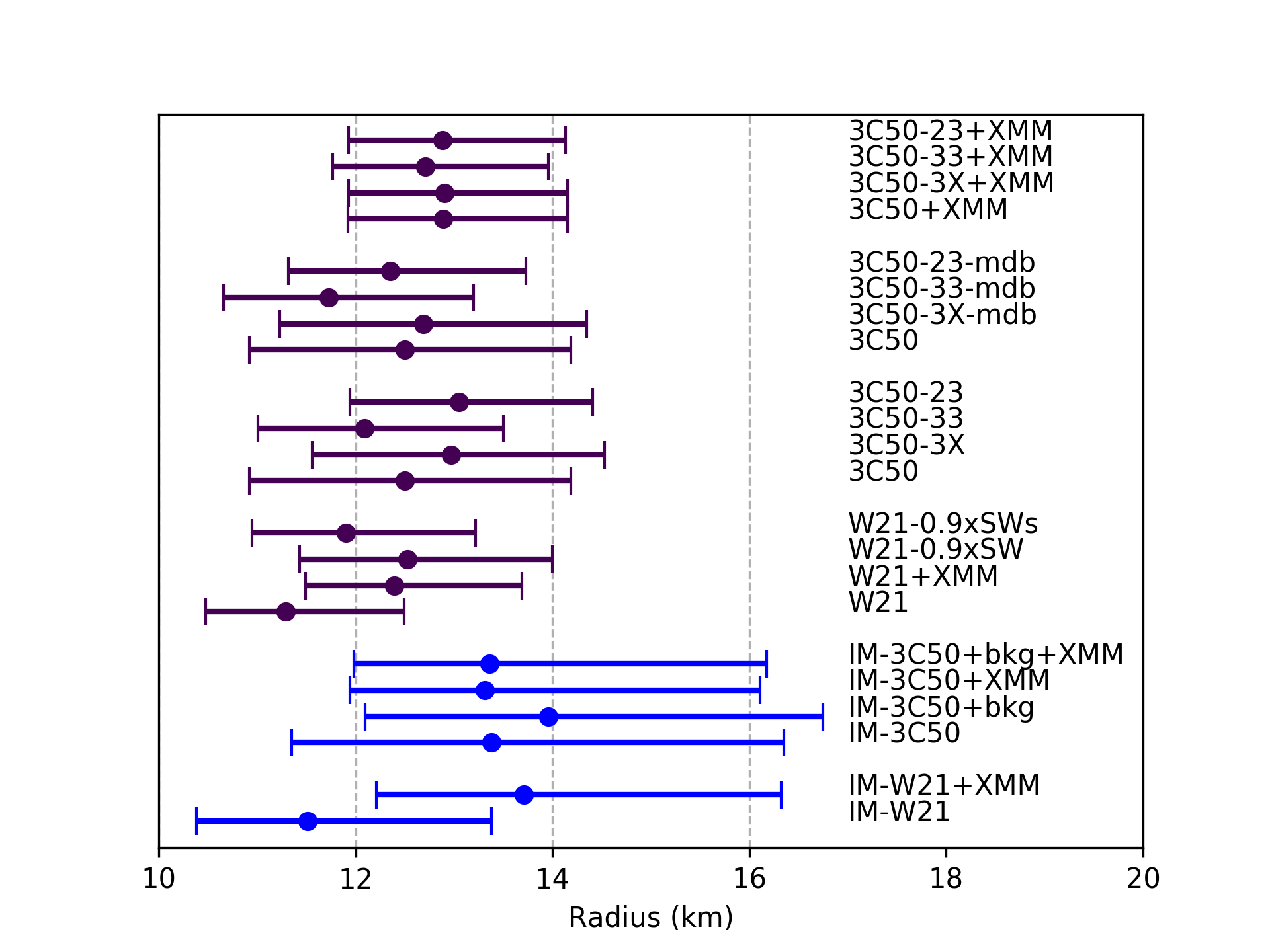}
    \caption{\small{
    Summary of the medians and $68.3\%$  credible intervals for radius using two circular hot spot model \texttt{ST-U} with different data sets and treatments for the background modeling. 
    Purple dots and lines correspond to the results obtained using \XPSI with models described in
    Tables \ref{table:sw}-\ref{table:3c50_XMM} and the associated figures in Section \ref{sec:results}. 
    Blue dots and lines correspond to the results obtained using the independent Illinois-Maryland (IM) code with the models described in Appendix \ref{apndx:A}, Table \ref{tab:NICERonly} and Table \ref{tab:NICERXMM}. 
    }}
    \label{fig:radius_intervals}
    \end{figure*}
}


\end{document}